\documentclass{svjour3}                     
\smartqed  
\usepackage{subfigure}
\usepackage{graphicx}
\usepackage{mathptmx}      
\usepackage{natbib}
\usepackage{aas_macros}
\journalname{Exp Astron}
\begin{document}

\title{POLARIX: a pathfinder mission of X-ray polarimetry
}


\author{Enrico Costa \and Ronaldo Bellazzini \and Gianpiero Tagliaferri
\and Giorgio Matt \and Andrea Argan \and Primo Attina' \and Luca
Baldini \and Stefano Basso \and Alessandro Brez \and Oberto
Citterio \and Sergio Di Cosimo  \and Vincenzo Cotroneo \and Sergio
Fabiani \and Marco Feroci \and Antonella Ferri \and Luca Latronico
\and Francesco Lazzarotto \and Massimo Minuti \and Ennio Morelli
\and Fabio Muleri \and Lucio Nicolini \and Giovanni Pareschi \and
Giuseppe Di Persio\and Michele Pinchera \and Massimiliano Razzano
\and Luigia Reboa \and Alda Rubini  \and Antonio Maria Salonico
\and Carmelo Sgro' \and Paolo Soffitta \and Gloria Spandre \and
Daniele Spiga \and Alessio Trois}


\institute{Enrico Costa  \and Andrea Argan \and Sergio Di Cosimo
 \and Sergio Fabiani \and Marco Feroci \and
Francesco Lazzarotto \and Ennio Morelli \and Fabio Muleri \and
Giuseppe Di Persio \and Paolo Soffitta \and Alessio Trois \at
Istituto di Astrofisica Spaziale e Fisica Cosmica, Via del Fosso
del Cavaliere 100, I-00133 Rome, Italy \\
\email{enrico.costa@iasf-roma.inaf.it} \and Ronaldo Bellazzini
\and Luca Baldini \and Alessando Brez \and Luca Latronico \and
Massimo Minuti \and Michele Pinchera \and Massimiliano Razzano
\and Carmelo Sgro' \and Gloria Spandre \at Istituto Nazionale di
Fisica Nucleare, Largo B. Pontecorvo 3, I-56127 Pisa, Italy \and
Giampiero Tagliaferri \and Stefano Basso \and Oberto Citterio \and
Giovanni Pareschi \and Daniele Spiga \at Osservatorio Astronomico
di Brera, Via E. Bianchi 46, I-23807 Merate (LC), Italy \and
Giorgio Matt \at Dipartimento di Fisica "E. Amaldi", Universita'
degli Studi Roma Tre, Via della Vasca Navale 84, 00146 Rome, Italy
\and Primo Attin$\grave{a}$ \and Antonella Ferri \at Thales Alenia
Space-Italia s.p.a., Strada Antica di Collegno, 253, I-10146,
Turin, Italy \and Antonio Maria Salonico \and Luigia Reboa \at
Telespazio, Via Tiburtina 965, I-00156 Rome, Italy \and Lucio
Nicolini \at Thales Alenia Space-Italia s.p.a., Strada Statale
Padana Superiore 290, I-20090 Vimodrone (Mi), Italy}

\date{Received: date / Accepted: date}

\maketitle

\begin{abstract}
Since the birth of X-ray astronomy, spectral, spatial and timing observation
improved dramatically, procuring a wealth of information on the majority of the
classes of the celestial sources. Polarimetry, instead, remained basically
unprobed. X-ray polarimetry promises to provide additional information procuring
two new observable quantities, the degree and the angle of polarization.
Polarization from celestial X-ray sources may derive from emission mechanisms
themselves such as cyclotron, synchrotron and non-thermal bremsstrahlung, from
scattering in aspheric accreting plasmas, such as disks, blobs and columns and from
the presence of extreme magnetic field by means of vacuum polarization and
birefringence. Matter in strong gravity fields and Quantum  Gravity effects can
be studied by X-ray polarimetry, too.

POLARIX  is a mission dedicated to X-ray polarimetry. It exploits
the  polarimetric response of a Gas Pixel Detector, combined
with position sensitivity, that, at the focus of a telescope,
results in a huge increase of sensitivity. The heart of the
detector is an Application-Specific Integrated Circuit (ASIC)
chip with 105600  pixels each one containing a
full complete electronic chain to image the track produced by the
photoelectron. Three Gas Pixel Detectors are coupled with three X-ray optics which
are the heritage of JET-X mission.
A filter wheel hosting calibration sources
unpolarized and polarized is dedicated to each detector for
periodic on-ground and in-flight calibration.
POLARIX will measure time resolved X-ray polarization
with an angular resolution
of about 20 arcsec in a field of view of 15 arcmin $\times$ 15 arcmin
and with an energy resolution of 20 $\%$ at 6 keV.
The Minimum Detectable Polarization is  12$\%$
for a source having a flux of 1 mCrab  and 10$^{5}$ s
of observing time.

The satellite will be placed in  an equatorial orbit of 505 km of
altitude by a Vega launcher.  The telemetry down-link station will
be Malindi.
The pointing of POLARIX satellite will be
gyroless and it will perform a double pointing during the earth occultation
of one source, so maximizing the scientific return.
POLARIX data are for
75 $\%$ open to the community while 25 $\%$ + SVP (Science Verification Phase,
1 month of operation) is dedicated to a core program activity
open to the contribution of associated scientists. The
planned duration of the mission is one year plus three months of
commissioning  and SVP, suitable to perform
most of the basic science within the reach of this instrument.
A \textit{nice to have} idea is to use the same existing mandrels
to build two additional telescopes of iridium with carbon coating plus two more
detectors. The effective area in this case would be almost
doubled.

\end{abstract}

\keywords{X-ray polarimetry \and Satellite missions}

\section{Introduction} \label{sec:Intro}
Since the early age of X-ray Astronomy, polarimetry
has been suggested as a powerful tool for a better understanding
of the physics and geometry of celestial sources. Unfortunately X-ray
polarimetry is still to be developed despite its scientific importance.

Non thermal processes play a major role in most subtopics of X-ray
Astronomy. Moreover the energy transfer in the inner regions of
compact X-ray sources is based on the interaction of radiation
with matter that highly deviates from spherical symmetry. Last but
not least the radiation, in its path to the observer crosses
regions of extremely high magnetic field that can produce
birefringence and/or extreme and very variable gravitational fields
that can deviate the radiation itself by effects of General
Relativity. The traditional
methods of Bragg diffraction around 45$^{\circ}$ and Compton/Thomson
scattering around 90$^{\circ}$ were affected either by the poor
efficiency or by the high background and/or by the large
systematic effects. Moreover both methods required the rotation of
the whole detecting apparatus or of a large part of it. This was
not a problem in the beginning because the whole X-ray Astronomy
was performed with slat or modulation collimators and the rotation
of satellites, around the unique stabilization axis, was the
baseline measurement technique for any mission.
The introduction of X-ray optics, firstly with the Einstein mission, produced a dramatic
improvement in the sensitivity of X-ray Astronomy disclosing the
possibility of deep extragalactic studies. The use of the X-ray optics removed the
need to rotate the satellite, therefore polarimetry based on the classical
techniques, that require rotation, became seriously mismatched with
imaging and spectroscopy. These classical techniques were, actually, based on
instruments that would be a major complication in spacecraft
design. As a result no polarimeters were included in major X-ray
missions by NASA or ESA. The strength of the science case was
not convincing enough to reach the decision for a dedicated mission
that, in any case, would have covered, a very limited sample
of bright sources.
The effort in detecting polarization from celestial X-ray sources
resulted, with the classical techniques,
in the unique positive measurement of the Crab Nebula. The first
measurement was done with a Bragg-polarimeter at 2.6 keV and
5.2 keV on board of a sounding rocket(Novick R., et al, 1972 \cite{Novick1972}).
A more accurate measurement was performed at the same energies, with a Bragg-polarimeter,
on-board OSO-8 satellite (Weisskopf, M. et al 1976, \cite{Weisskopf1976}) also excluding
the contribution of the pulsar by means of lunar occultation technique
(Weisskopf, M. et al, 1978 \cite{Weisskopf1978}). The degree of polarization
from the nebula that was found is 19.2 $\%$ $\pm$ 1.0$\%$.
The Bragg polarimeter on-board OSO-8 provided coarse upper limits on
many celestial X-ray sources (Hughes, J.P. et al, 1984 \cite{Hughes1984})
and an accurate zero-measurement on Sco X-1 (Long, K.S., et al 1979 \cite{Long1979}).
Compton gamma-ray polarimetry resulted in the very debated measurement of some bright
gamma-ray bursts(\cite{CoburnBoggs2003}\cite{Wigger2005}), and in the measurement
of the polarization of the Crab (100 keV-1 MeV) by means of instruments not
specifically designed for polarimetry on board of INTEGRAL satellite
(\cite{Dean2008},\cite{Forot2008}).

The development of detectors based on the photoelectric effect
that can measure simultaneously the interaction point, the energy,
the arrival time of the photon, together with the emission angle
of the photoelectron, opens the possibility to perform focal plane
polarimetry, namely to introduce in this subtopic of X-ray
Astronomy the same jump in sensitivity of the other techniques.
The meagerness of data on polarization of sources is such that
also a small mission, namely with a telescope set of few hundreds
square centimeters, can allow for the measurement of polarization
on tens of sources opening this window in the violent sky of
X-ray Astronomy. Focal plane polarimetry requires anyway a
significant amount of photons. In any case the measurement will be
source dominated. Therefore the sensitivity is only a matter of
total area whether or not these photons are collected by a
single telescope or by a cluster of telescopes. A cluster of
telescopes with a focal length compatible with small launchers can
be a solution for an ambitious pathfinder within a limited budget.
The passage for such a pathfinder is highly desirable to reach
a first assessment of the discipline and to better adjust the
design of polarimeters to be included in future large telescopes
such as XEUS/IXO.
In Italy we are fortunate to have developed imaging
polarimeters (\cite{Costa2001}\cite{Bellazzini2006}) to a high
degree of readiness and to have three X-ray telescopes
of excellent quality available, remaining from the JET-X experiment aboard
the never flown SPECTRUM-X-Gamma mission. POLARIX is designed to
combine these instruments. It has been proposed as a Small Mission
to  be launched with a VEGA rocket in response to an Announcement
of opportunity for Small Scientific Missions issued by ASI in
2007. POLARIX was selected to perform a phase A study. In this
paper we report the advanced design deriving from this study. The
selection by ASI of two possible missions that are to be flown is in
progress.

\section{Scientific objectives}
We describe here the main scientific objectives of POLARIX, as can
be derived from the present literature. It is important to note,
however, that X-ray polarimetry is an almost unexplored field
therefore there is large room for unexpected discoveries and for
further theoretical analysis.

The discussion is divided in two sections, where we describe how
X-ray polarimetry can help in understanding the Astrophysics of
cosmic sources, and, respectively, how astrophysical sources can
be used as laboratories to test theories of Fundamental Physics.

In the following, all numerical examples concerning Minimum Detectable Polarization
(MDP, defined in $\oint 7.2$) will be based on the configuration of
three telescopes. The MDP for the configuration of five telescopes
can be approximately obtained after multiplication
by a factor $\sqrt{2}$. The two additional telescopes almost double the
original area because of carbon coating (as discussed below).

\subsection{Astrophysics}

\subsubsection{Acceleration Phenomena}
Acceleration of particles is the clue of Cosmic Ray Physics. Super
Nova Remnants (both plerionic and shell-like) are the best
candidates for the acceleration of the bulk of the electrons
reaching the Earth. Ultra High Energy Cosmic Rays are instead very
likely to be originated outside the Galaxy, most probably in jetted AGN,
as shown by recent Auger results (Abraham et al.
2007,\cite{Abraham2007}). X-ray polarimetry is a powerful probe to
investigate acceleration phenomena, since energetic particles in a
magnetic field emits synchrotron radiation, which is highly
polarized.

\subsubsection{Acceleration Phenomena : Pulsar Wind Nebulae}
\begin{figure}
\centering
\includegraphics [scale=0.30]{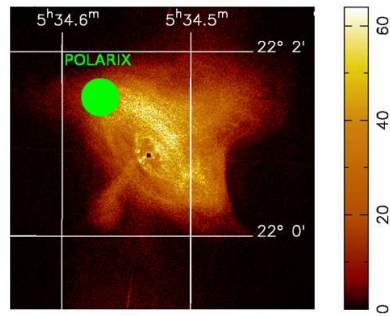}
\caption{The Crab Nebula as observed by Chandra (Weisskopf et al.,
2000\cite{Weisskopf2000}). The green circle is
the POLARIX point spread function}
\label{fig:1}       
\end{figure}

The Crab Nebula is one of the best studied astronomical sources.
The first, and the only one so far, polarimetric measurement in the
X-rays showed that this source is highly polarized, proving
synchrotron as the emission mechanism. The pulsar itself is also
highly polarized in the radio to optical bands. Recent INTEGRAL
results (Dean et al. 2008\cite{Dean2008}, Forot et al.
2008\cite{Forot2008}) showed high level of polarization  also in
gamma-rays, with the polarization angle aligned to the optical one
(and to the rotation axis of the pulsar). The X-ray emission of
the nebula is highly structured, as shown by the wonderful Chandra
images, with a torus-plus-jet geometry. Detailed and
space-resolved measurements by POLARIX (fig.\ref{fig:1}) will
allow us to study the dynamics of the plasma in the nebula and the
acceleration mechanism of the pulsar, which is able to convert the
large part of its rotational energy in accelerating particles
which eventually shine in the nebula by synchrotron emission.

\subsubsection{Acceleration Phenomena : $\mu$QSO}
One of the challenges of the present day high energy astrophysics
is to understand how matter is accelerated in jets, and how the
mechanism responsible for the emission can work both in galactic
and extragalactic sources.

\begin{figure}
\centering
\includegraphics [scale=0.2]{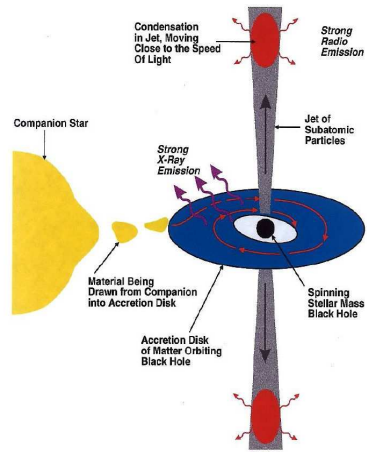}
\caption{A sketch of a microquasar}
\label{fig:2}       
\end{figure}

Regarding galactic sources, at present about 2 dozens X-ray
binaries show superluminal behavior radio-emitting spots moving
away from a compact core apparently (fig.\ref{fig:2}) faster than
the speed of light (Mirabel $\and$ Rodriguez 1994
\cite{Mirabel1994}).  These objects can be extremely luminous in
X-rays (10$^{38-39}$ erg s$^{-1}$ ; Belloni et al.
1997\cite{Belloni1997}). Their high
luminosities and high masses - inferred from optical determination
of the orbital parameters (e.g., Orosz $\and$ Bailyn
1997\cite{Orosz1997}) - indicate that the compact object is a
black hole in many of them. Their structure - black hole,
accretion disk, and relativistic jet - and their multiwavelength
behavior, including gamma-ray emission (e.g., Aharonian et al.
2006\cite{Aharonian2006}) and radio and optical polarization
(Nagae et al. 2008\cite{Nagae2008}, and references therein)
strongly resemble those of radio-loud quasars, of which they seem
to be just a scaled down version. For this reason they have been
named "microquasars". The study of their polarization properties
can help shading light on jet formation and its relation to
accretion, and the site  (disk, corona, or jet) of its origin,
which may also be applicable to AGN (e.g. Mirabel
2007\cite{Mirabel2007}), with the additional bonus of allowing us,
thanks to the much smaller time scales, to study their behavior
over a wide interval of the accretion rate.

These sources are also good candidates to search for General
Relativity effects which modify polarization properties, as will
be described in the "Fundamental Physics" section.

\subsubsection{Acceleration Phenomena : Blazar and Radiogalaxy}
AGN are customarily divided into two subclasses. i.e. radio-loud and radio-quiet AGN,
depending on the level of radio emission. In radio-loud AGN, a relativistic, highly collimated
jet is present; in the subclass of Blazars, it is directed close to the line-of-sight and all
Special Relativity effects are magnified. In particular, due  to Doppler boosting, jets dominate
the emission at all wavelengths.

\begin{figure}
\centering
\includegraphics[scale=0.60] {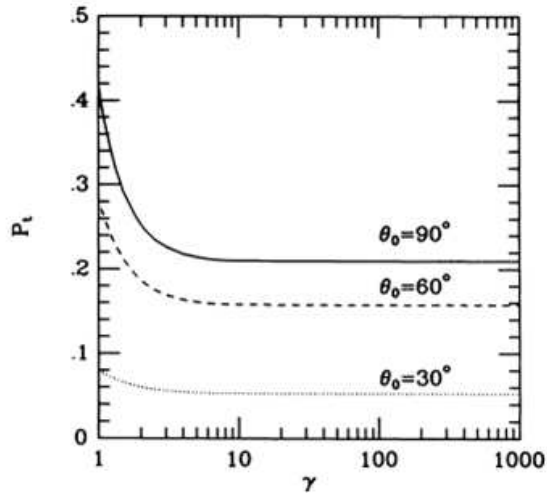}
\caption{The polarization degree of the SSC emission as a function of
the Lorentz factor of the electrons $\theta$$_{o}$ is the angle between the
observer and the magnetic field (from
Celotti $\&$ Matt(1994) \cite{Celotti1994}.)}
\label{fig:3}       
\end{figure}

The spectral energy distribution (SED) of Blazars is composed of
two peaks, the first one due to synchrotron emission, the second
one due to inverse Compton scattering (IC) of either the
synchrotron photons (synchrotron self-Compton, SSC) or external
photons, presumably from the accretion disk. In some cases, the
synchrotron peak dominates in X-rays, and strong polarization is
therefore expected. Comparison with the polarization in other
bands can elucidate on the structure of the jet. In other sources the
IC peak is instead observed in X-rays. In the latter case, X-ray
polarimetry will offer a simple way to establish if IC occurs on
synchrotron or external photons. While the polarization angles of
synchrotron and SSC emission are expected to be the same, and
perpendicular to the magnetic field (Celotti $\&$ Matt
1994\cite{Celotti1994}), in the external photons model the IC
polarization is related to the jet axis (Begelman $\&$ Sikora
1987\cite{Begelman1987}), and the polarization angle in the two
peaks needs no longer to be the same. In both models, the
polarization degree (fig.\ref{fig:3}) is expected to be very high,
up to 50 $\%$ or more unless the electrons responsible for the IC
emission are hot (see also Poutanen 1994\cite{Poutanen1994}).
Multiwavelength polarimetry will therefore provide unique
information on the emission mechanism.

In non-Blazar radio-loud AGN, the jet is directed away from the
line-of-sight, and the jet emission no longer dominates over the
disk-related emission. Interestingly, for a few bright sources
(most notably the famous bright quasar 3C 273), Grandi $\&$
Palumbo (2004\cite{Grandi2004}, 2007\cite{Grandi2007}) have
suggested that the two components are of comparable importance in
the 2-10 keV band. These results are based on a spectral
deconvolution, and therefore on assumptions on the spectral shapes
of the various components. X-ray polarimetry offers a
model-independent way to test this hypothesis: as the two
components have different spectra (the jet spectrum being harder),
and are polarized with different polarization angles, a rotation
of the angle with energy is expected, in accordance with the
energy-dependent weights of the two components.

\subsubsection{Emission in strong magnetic fields}
Ordered magnetic fields cause radiation to be polarized not only
because of synchrotron emission (as in the acceleration phenomena
discussed above) but also, if strong enough, because they channel
the matter along the flux lines, resulting in strong asphericities
in the matter distribution. Moreover, plasma opacity in a strong
magnetic field is different in the two modes, leading to strong
polarization of the emerging radiation.

\subsubsection{Emission in strong magnetic fields: Magnetic Cataclysmic Variables}
\begin{figure}[htpb]
\centering
\subfigure[\label{fig.4a_}]{\includegraphics[scale=0.15]{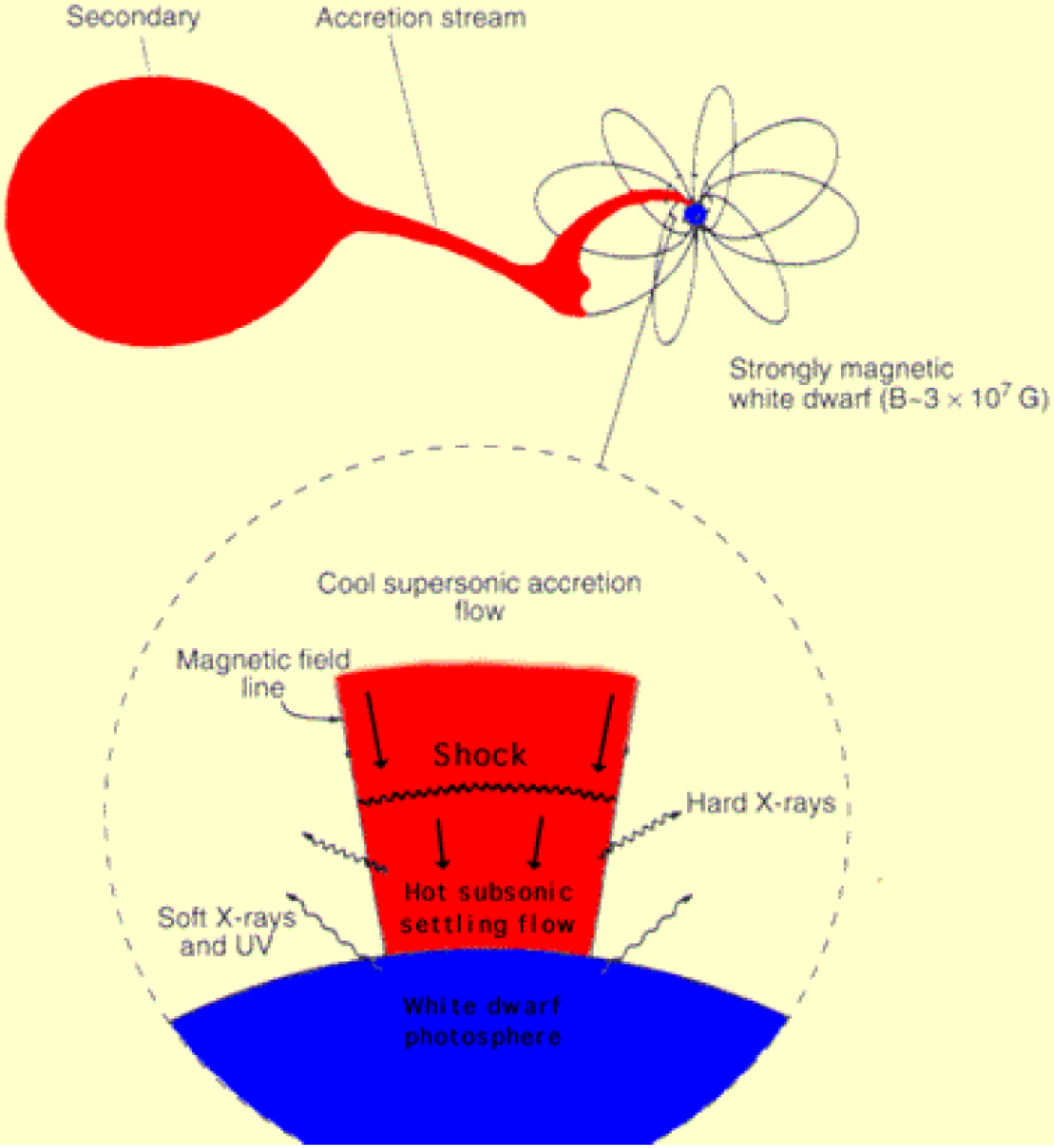}}
\hspace{1cm}
\subfigure[\label{fig.4b}]{\includegraphics[scale=0.5]{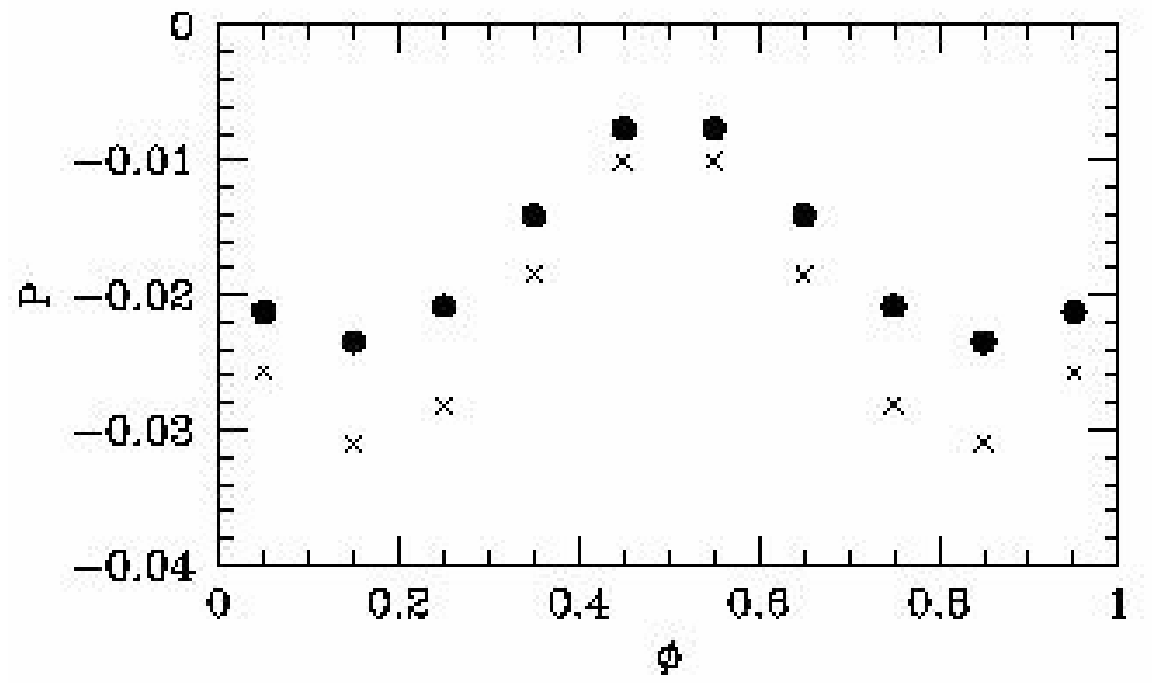}}
\caption{({\bf a}). Sketch of the accretion column in a magnetic CV. ({\bf b}). The phase dependence
 of the polarization degree expected in \textit{AM Her} in two energy bins (5-10 keV), filled circles,
  (10-15 keV), crosses. From Matt (2004)\cite{Matt2004}.}
\end{figure}

Magnetic cataclysmic variables (mCVs), which include polars and
intermediate polars, are binaries with a strongly magnetized white
dwarf (WD) accreting material (see fig.\ref{fig.4a_}) from a
Roche-lobe filling low-mass stars (see Warner
1995\cite{Warner1995} for a comprehensive review). The magnetic
field is strong enough (1-100 Mgauss) to channel the accretion
flow directly to the WD, preventing the formation of an accretion
disk in polars and magnetically truncating the disk in
intermediate polars (which have a slightly lower magnetic field,
but are generally stronger X-rays emitters, because of higher
accretion rates). The accreting matter is heated to keV
temperatures in a standing shock near the WD surface. The
post-shock material is cooled by emitting optical-IR cyclotron
radiation and bremsstrahlung in X-rays. The X-rays  are in part
scattered and reflected by the WD surface, the disk (if present)
and the magnetosphere. The polarization characteristics are shown
in fig. \ref{fig.4b}

Even if the X-ray emission is mainly due to bremsstrahlung,
scattering opacity in the accretion column could not be negligible
for high accretion rates, and the emission may be polarized. The
polarization depends on the viewing inclination of the system and
is sensitive to the system configuration. The viewing orientation of
the accretion column varies with the orbital period, and the
polarization signals is therefore periodic, with an amplitude
reaching 4$\%$ - 8$\%$ for axisymmetric models (Matt
2004\cite{Matt2004}, McNamara et al. 2008 \cite{McNamara2008}).
Reflection from the WD surface, which is relevant above a few keV,
is also expected to be significantly polarized, providing a
characteristic energy dependence of the polarization properties
(Matt 2004\cite{Matt2004}). Several bright intermediate polars,
and certainly the brightest polar, AM Her, when in high state, can
be searched for phase-dependent polarization with POLARIX.

\subsubsection{Emission in strong magnetic fields: Accreting millisecond pulsars}
\begin{figure}
\centering
\includegraphics[scale=0.40] {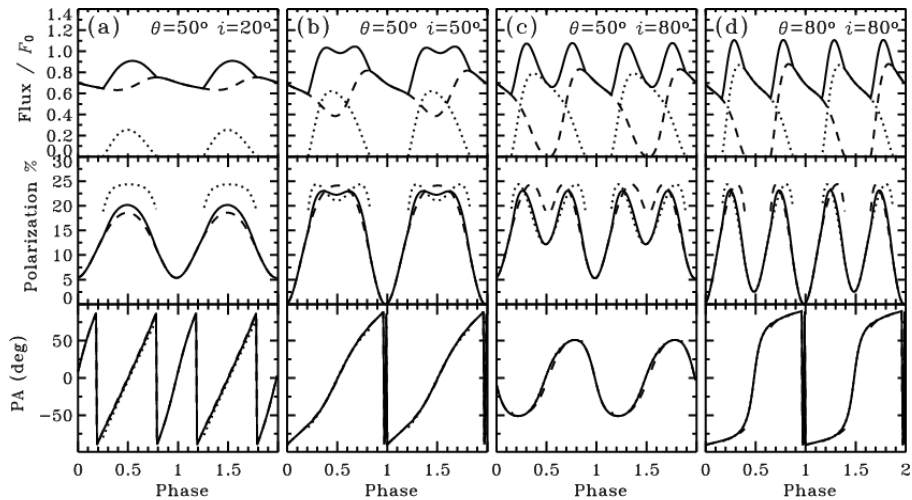}
\caption{Lightcurves of the flux, polarization
degrees and angle in an accreting millisecond pulsar
for different set of geometrical parameters. From Viironen $\&$ Poutanen(2004) \cite{Viironen2004}}
\label{fig:5}       
\end{figure}

Accretion in close binary systems can spin up the neutron-star
rotation, resulting in accretion-powered millisecond pulsars
(aMSPs). At present, eight sources are known, with periods ranging
from 1.67 to 5.49 ms.  Their spectra consist of a blackbody
component, likely originating in a hot spot on the neutron star
surface and with typical temperatures of about 1 keV, and a hard
power-law component, probably due to Comptonization in a radiative
shock surface, with a temperature of 30 - 60 keV and  optical
depths $\sim$ 1-2. The observed pulsations indicate that the shock
covers only a small part of the neutron-star surface. The
scattered radiation should be linearly polarized
(fig.\ref{fig:5}), with the polarization degree and angle varying
with the phase (Viironen $\&$ Poutanen 2004\cite{Viironen2004}).

\subsubsection{Emission in strong magnetic fields: Accreting X-ray pulsars}
\begin{figure}[htpb]
\centering
\subfigure[\label{fig:6a_}]{\includegraphics[scale=0.5]{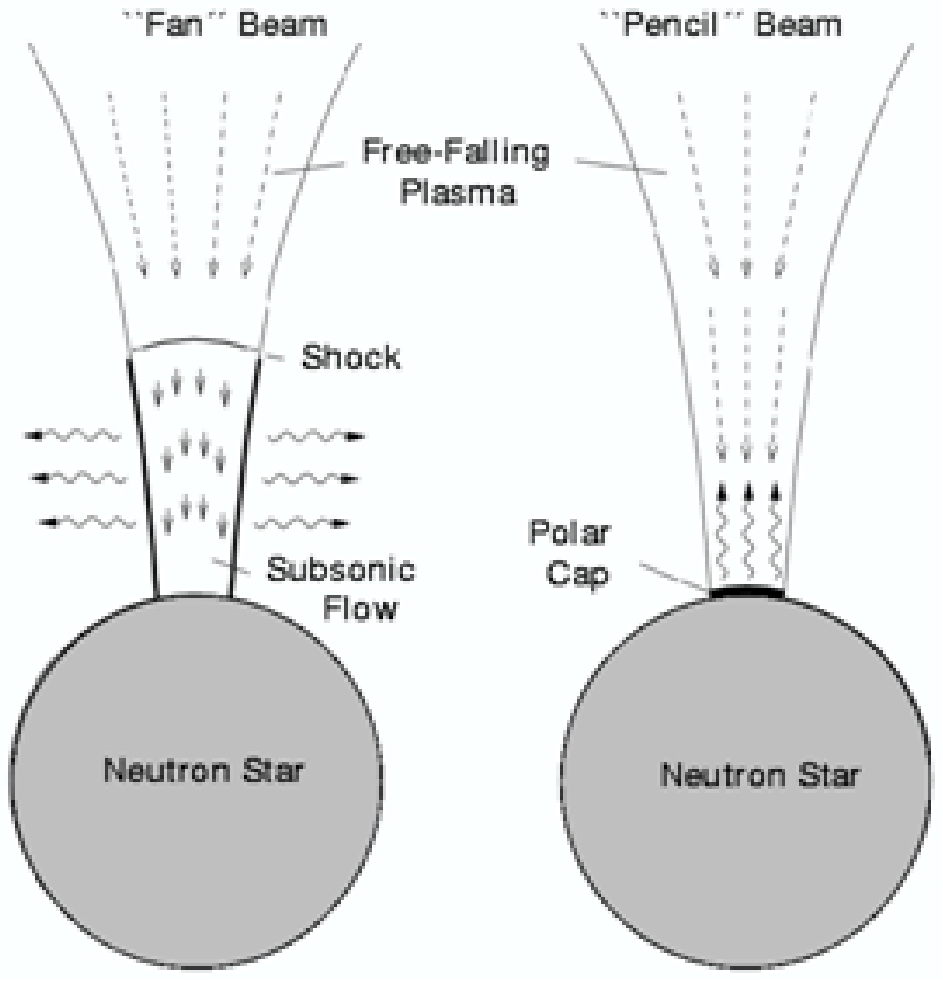}}
\hspace{0.3cm}
\subfigure[\label{fig:6b}]{\includegraphics[scale=0.25]{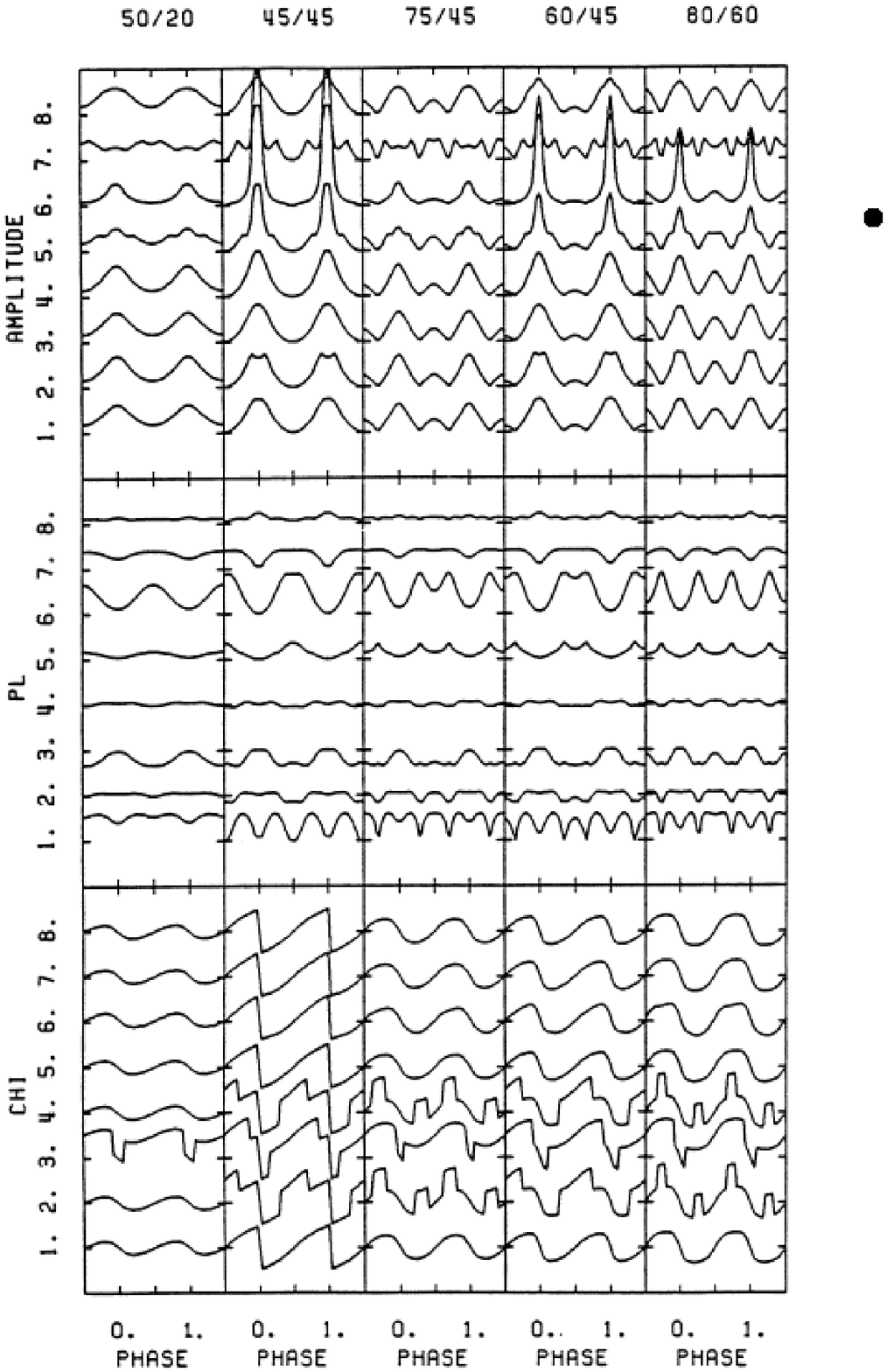}}
\caption{({\bf a}). Sketch of an X-ray pulsar in the 'Fan' beam and 'Pencil'
 beam cases. ({\bf b}). Lightcurves of the flux, polarization degrees
 and angle in a X-ray pulsar for different sets of the geometrical parameters.
From Meszaros et al. (1988)\cite{Meszaros1988}}
\end{figure}

Accreting X-ray pulsars are binary systems (fig. \ref{fig:6a_}) in
which the compact object is a Neutron Star, with very strong
magnetic fields 10$^{12}$-10$^{13}$ G, as derived from the
detection of cyclotron lines. In such a strong a field, a
birefringence effect, due to the different plasma opacity to the
ordinary and extraordinary modes, arises, resulting in a strong
linear polarization of the emerging radiation.

Detailed calculations (e.g. M$\acute{e}$z$\acute{a}$ros et al.
1988 \cite{Meszaros1988}, fig. \ref{fig:6b}) show that the linear
polarization depends strongly on the geometry of the emission
region (accretion column), and varies with energy and pulse phase,
reaching very high degrees, up to 70$\%$ for favorable
orientations. The detailed properties of the X-ray spectrum, pulse
profile, and polarization depend on the assumptions made on the
physical and geometrical properties of the systems, but
nevertheless some general conclusions  can be reached. In
particular, phase-resolved polarimetry can distinguish between
"pencil" and "fan" radiation patterns, a long standing problem
still awaiting a firm solution. Because the degree of linear
polarization is maximum for emission perpendicular to the magnetic
field, the flux and degree of polarization are in-phase for fan
beams, but out-of-phase for pencil beams. In cases when pulse
profiles change dramatically with energy, it is possible that both
fan and pencil beam components are present, each component
dominating at different energies.

\subsubsection{Scattering in aspherical situations:X-ray Binaries}
Even when the magnetic field of the compact object is not strong
enough to channel the accreting matter, asphericities are present
because the matter usually forms accretion disks. GR effects are
expected to be important for the emission originated in the inner
regions of the disk, close to the black hole or the neutron star.
These effects will be discussed in the "Fundamental Physics"
section. Here we just emphasize that in accretion-disc-fed sources
the hard component (which is the dominant one in the 2-10 keV
spectrum where the sources are in the so called hard state) is
likely due to Comptonization of disk photons in a hot corona, and
it is therefore expected to be strongly polarized (e.g. Haardt
$\&$ Matt 1993 \cite{Haardt1993}, Poutanen and Vilhu 1993
\cite{Poutanen1993}). The polarization degree will put constraints
on the, so far unknown, geometry of the corona.

\begin{figure}[htpb]
\centering
\subfigure[\label{fig.7a_}]{\includegraphics[scale=0.45]{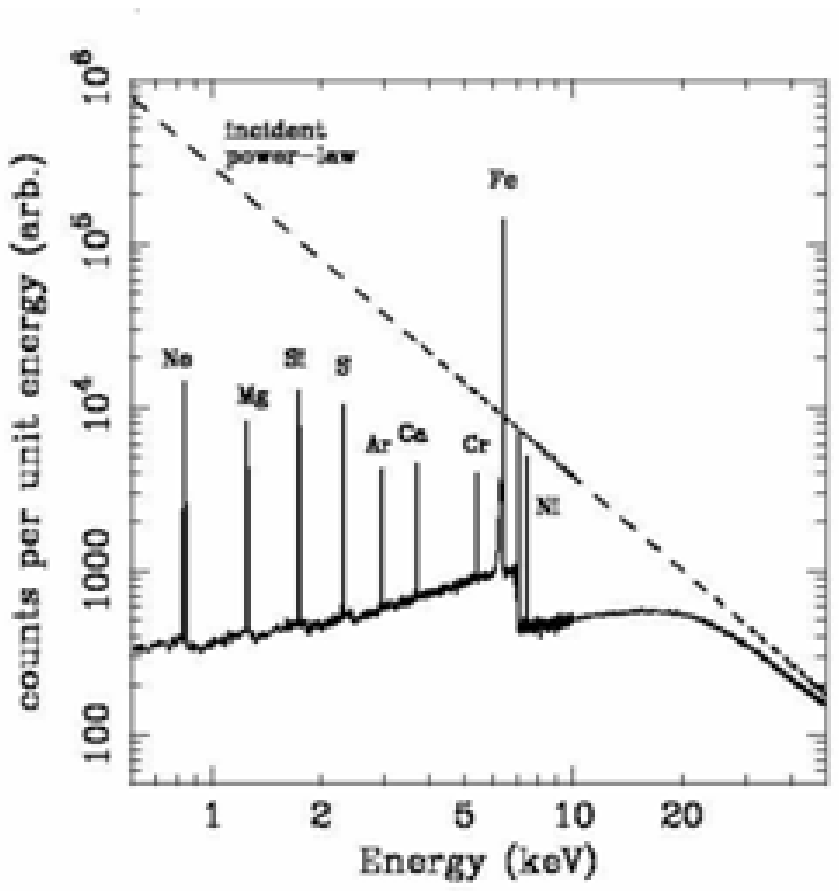}}
\hspace{1cm}
\subfigure[\label{fig.7b}]{\includegraphics[scale=0.5]{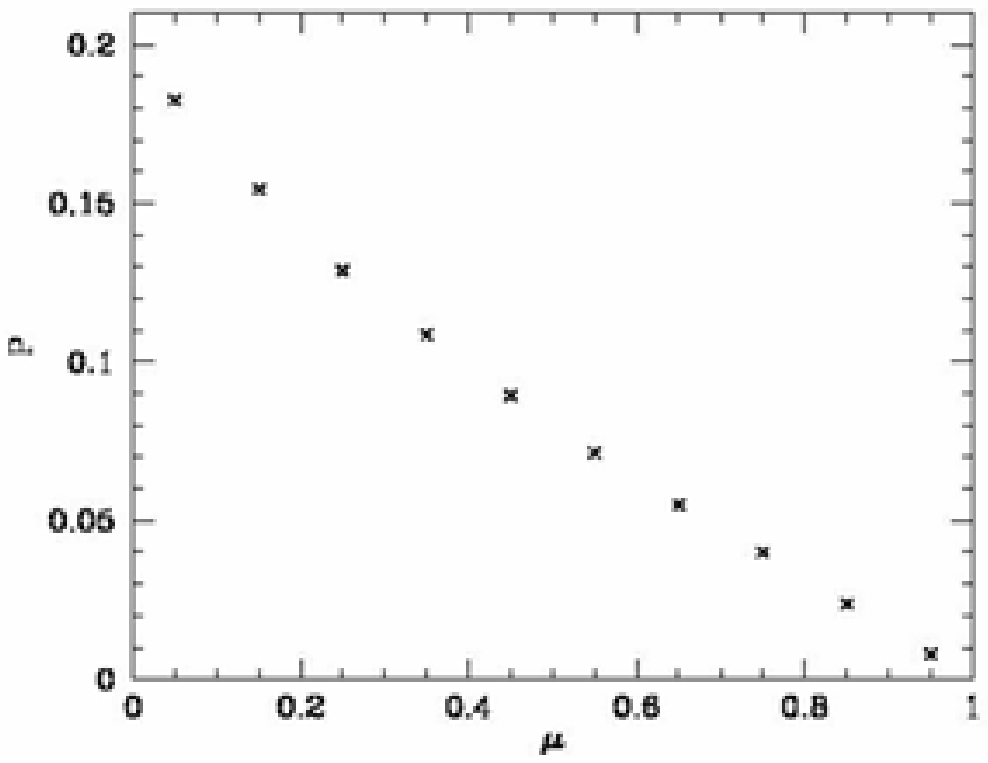}}
\caption{({\bf a}). The reflection spectrum from an externally illuminated
slab(Reynolds et al. 1995 ({\bf b}). The polarization degree of the reflected
radiation as a function of the cosine of the inclination angle of the slab
(Matt et al. 1989 \cite{Matt1989})}
\end{figure}

Part of the primary emission is intercepted and reflected by the
accretion disk itself, giving rise to the so-called Compton
Reflection (CR) component (fig.\ref{fig.7a_} . This component,
which becomes relevant above 7 keV or so, is also highly polarized
, the polarization degree depending mainly on the inclination
angle of the disk (fig.\ref{fig.7b}, Matt et al. 1989
\cite{Matt1989}).

\subsubsection{Scattering in aspherical situations:Radio-quiet AGN}
\begin{figure}
\centering
\includegraphics[scale=0.40] {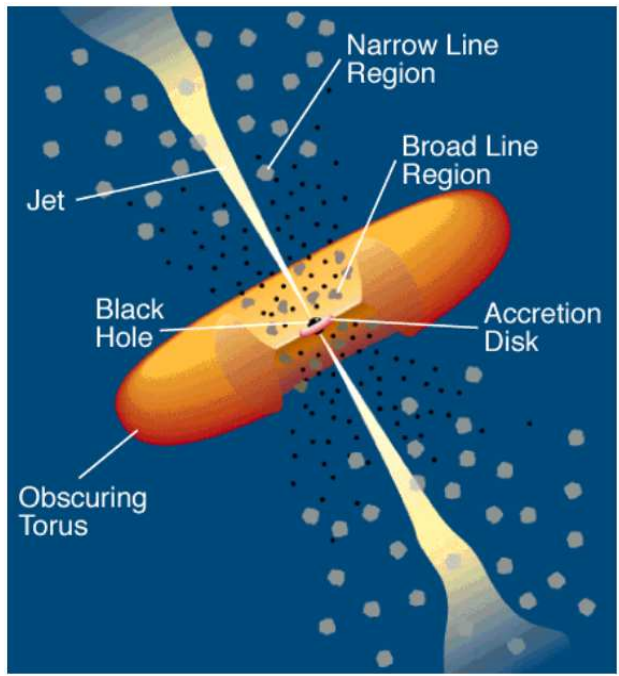}
\caption{The unification model of AGN: the classification of the source
depend on the orientation. In radio-loud AGN a jet is also present
while in radio quiet AGN is not. From Urry and Padovani (1995)\cite{Urry1995}.}

\label{fig:8}       
\end{figure}

The inner regions of radio-quiet AGN are just scaled-up versions
of those present in Galactic Black Hole systems (fig.\ref{fig:8}),
with the very important difference that here the Comptonization
component always dominates, as the disk thermal component is in
the UV/Soft X-ray band, due to the $T_{disk} \propto
M_{BH}^{-1/4}$ relation. For the Comptonization and CR
components, the same considerations made in the previous paragraph
still holds (but see the ''Fundamental Physics'' section for a
test of GR effects based on time variability).

In addition to the accretion disk, other reflecting regions are
present in radio-quiet AGN, first and foremost the so-called
torus envisaged in Unification models (Antonucci 1993
\cite{Antonucci1993}). Despite the name, the actual geometrical
shape of the 'torus' is basically unknown, and polarimetric
observations can help to solve this issue.

\subsubsection{Scattering in aspherical situations:X-ray Reflection Nebulae
 and the study of the Supermassive Black Hole in the  Galaxy}

\begin{figure}
\centering
\includegraphics[scale=0.5] {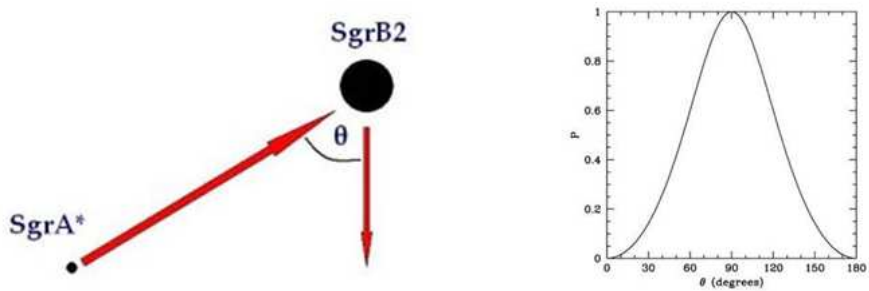}
\caption{In the right panel, the polarization degree of SgrB2 as a function of
the angle $\theta$ (defined in the left panel) is shown, assuming that the presently
SgrB2 is reflecting the radiation originated by the black hole at the
center of the Galaxy a few hundreds years ago}
\label{fig:9}       
\end{figure}

Sgr B2 is a molecular cloud in the Galactic Center region, located
at about 100 pc (projected distance) from Sgr A*, the supermassive
black hole in the center of the Galaxy. Its X-ray spectrum is well
reproduced by a pure Compton Reflection component, indicating that
Sgr B2 is reflecting the X-ray radiation produced by a source
outside the cloud (Sunayev et al. 1993\cite{Sunyaev1993}). The
puzzle here is that there is no X-ray source bright enough  in the
surroundings. It has therefore been proposed that SgrB2 is
reflecting past emission by the central black hole (Koyama et al.
1996\cite{Koyama1996}), which should therefore have undergone a
phase of activity about three hundreds years ago. If the  emission
from the nebula is indeed due to scattering, it should be
(Fig.\ref{fig:9}) very highly polarized (Churazov et al.
2002\cite{Churazov2002}), with a direction of polarization normal
to the scattering plane, and therefore to the line connecting Sgr
B2 to the illuminating source. Polarimetry will place a strong
limit on the position of the source which illuminated Sgr B2 in
the past and, if the direction of polarization will point as
expected towards Sgr A*, it will be proved that not many years ago
the Galaxy was a low luminosity AGN.
The precision with which  the polarization angle can be
measured depends on the polarization degree, but it is of the
order of a few degrees, good enough to set very tight constraints
on the origin of the illuminating radiation.

It must be noted that the flux from SgrB2 is varying with time (unfortunately
decreasing, Koyama et al. 2008\cite{Koyama2008}). However, other reflecting nebulae are present
around the central black hole, which are also varying with time
(Muno et al. 2007\cite{Muno2007}). The brightest of them when POLARIX will be on orbit will
of course be chosen for observation.

\subsection{Fundamental Physics}
X-ray polarimetry will also have an impact on fundamental physics, allowing for the study
of QED effects in extreme magnetic fields, of General Relativity effects in the strong
field regime, and even putting constraints on Quantum Gravity theories.

\subsubsection{Matter in Extreme Magnetic Fields}
\begin{figure}[htpb]
\centering
\subfigure[\label{fig.10a_}]{\includegraphics[scale=0.20]{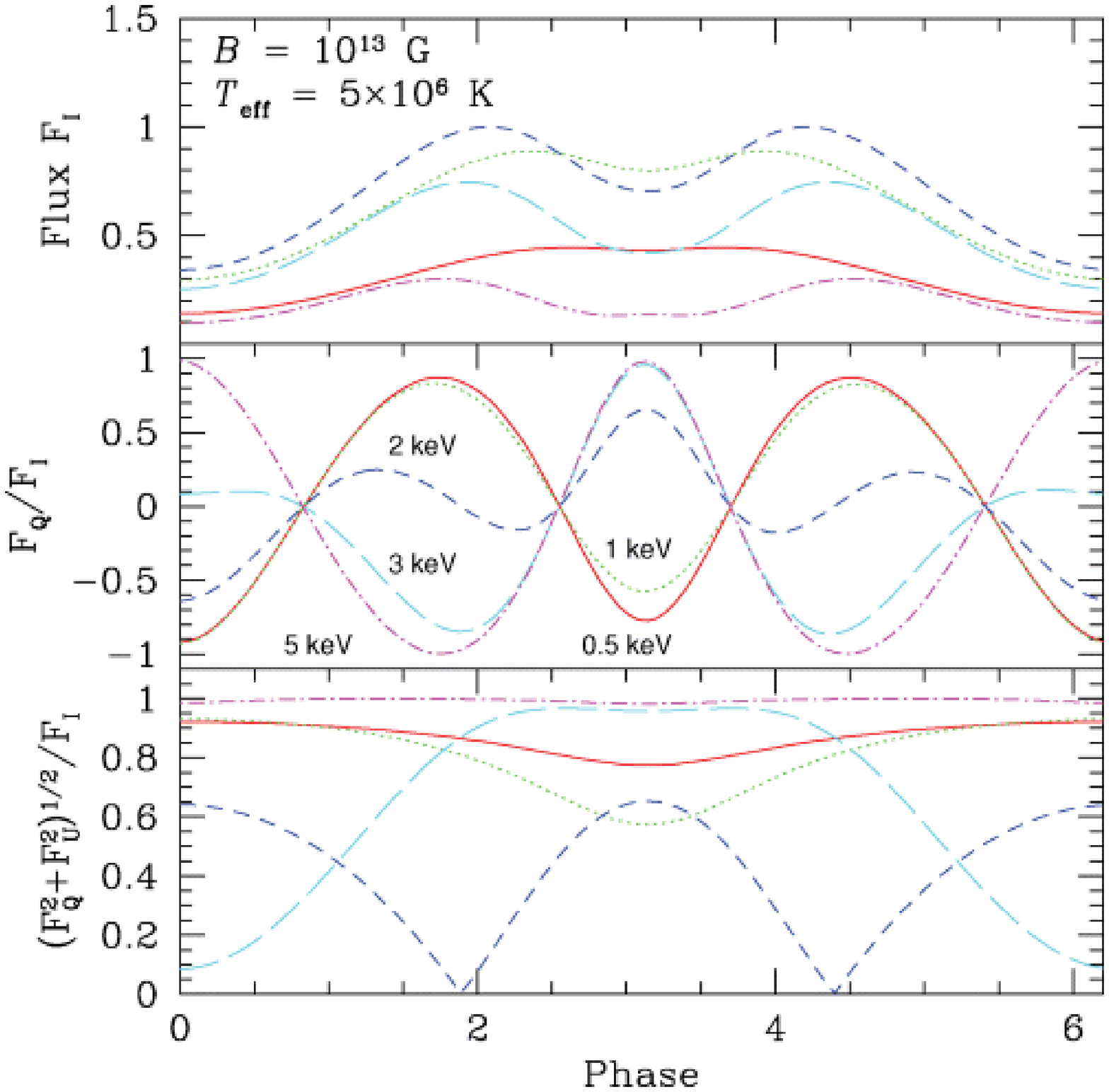}}
\hspace{0.2cm}
\subfigure[\label{fig.10b}]{\includegraphics[scale=0.20]{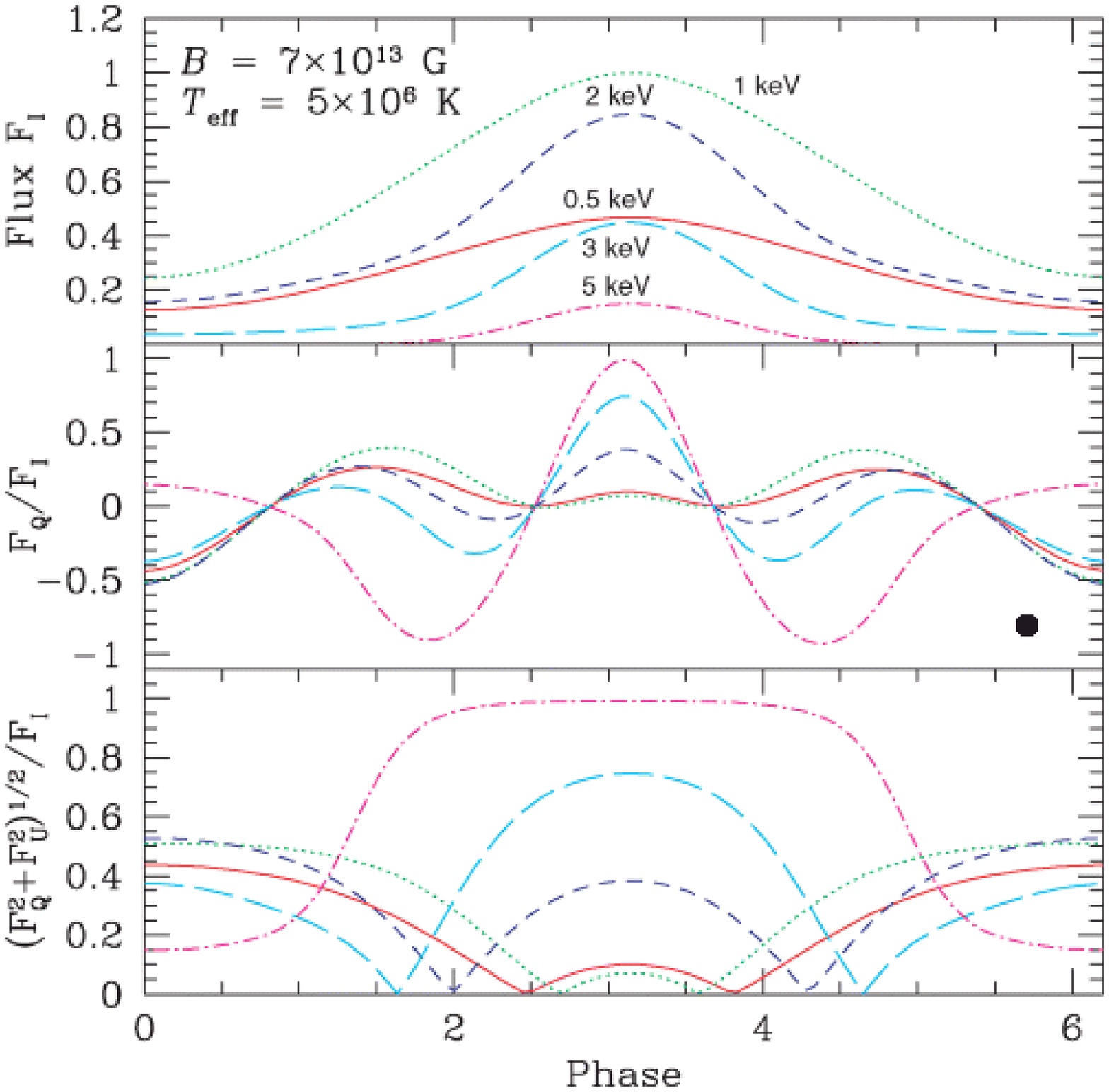}}
\hspace{0.2cm}
\subfigure[\label{fig.10c}]{\includegraphics[scale=0.20]{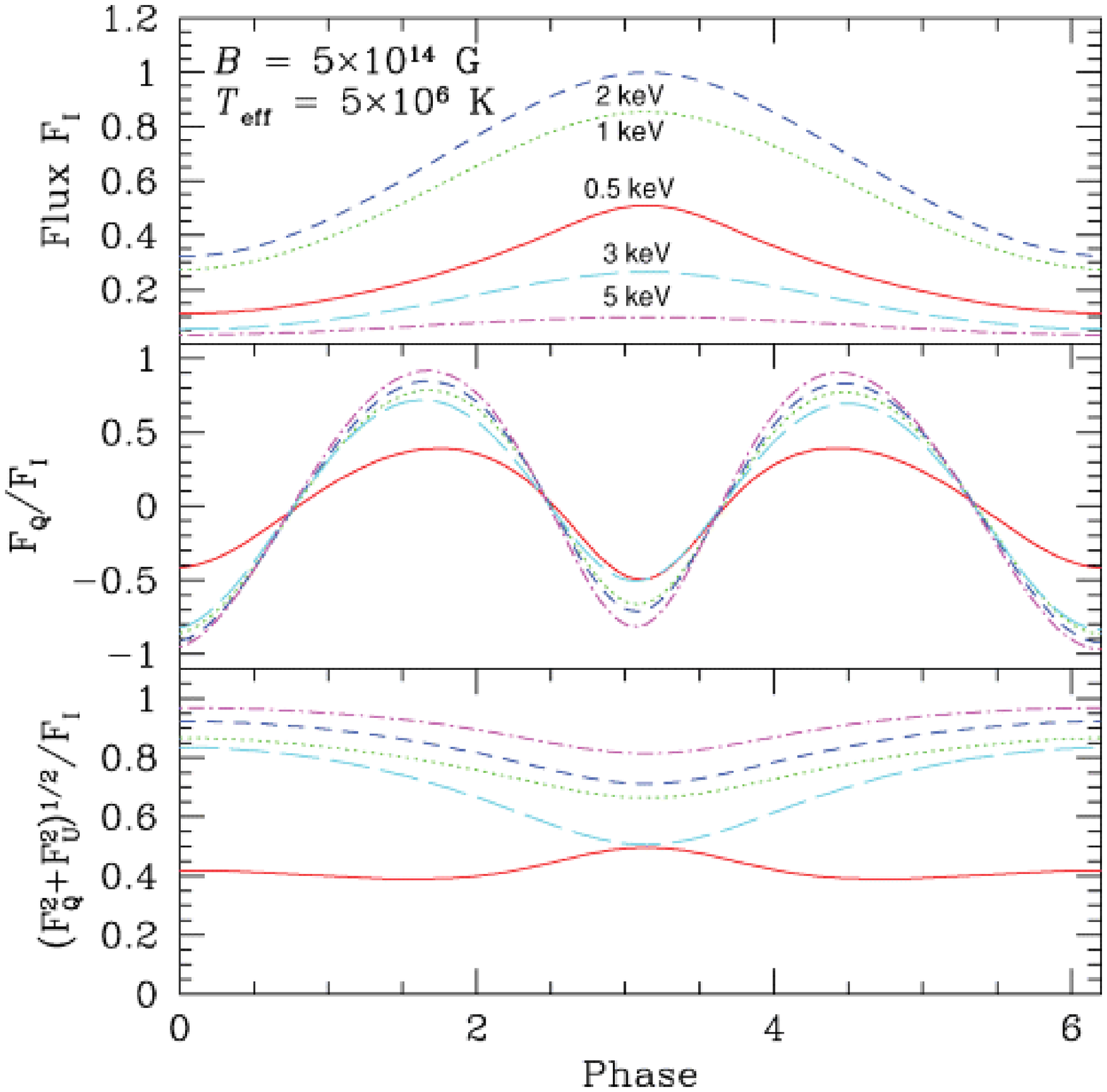}}
\caption{Light curves of the flux, polarization degree and angle
in an isolated star for different energies and,
in different panels, for different magnetic fields.
From van Adelsberg $\&$ Lai (2006)\cite{Adelsberg2006}}
\end{figure}

As mentioned above, X-ray polarimetry will also allow us to observe a
quantum-electrodynamic (QED) effect, i..e the vacuum birefringence
induced by a strong magnetic field. Predicted nearly 70 years ago
(Heisenberg $\&$ Euler 1936 \cite{Heisemberg1936}), the effect
could not be verified observationally so far. Detailed
calculations of this effect in isolated neutron stars (van
Adelsberg $\&$ Lai 2006 \cite{Adelsberg2006}) have shown a strong
energy dependence of the polarization pattern for a single source
(fig.\ref{fig.10a_}, fig.\ref{fig.10b},fig.\ref{fig.10c}) and a
strong B-dependence when different sources, with different
magnetic fields, are compared. Vacuum birefringence has a much
smaller impact on spectral parameters, leaving polarimetry as
by far the best tool to observe this effect.

\subsubsection{Matter in Extreme Gravitational Fields}
\begin{figure}[htpb]
\centering
\subfigure[\label{fig:11a}]{\includegraphics[scale=0.5]{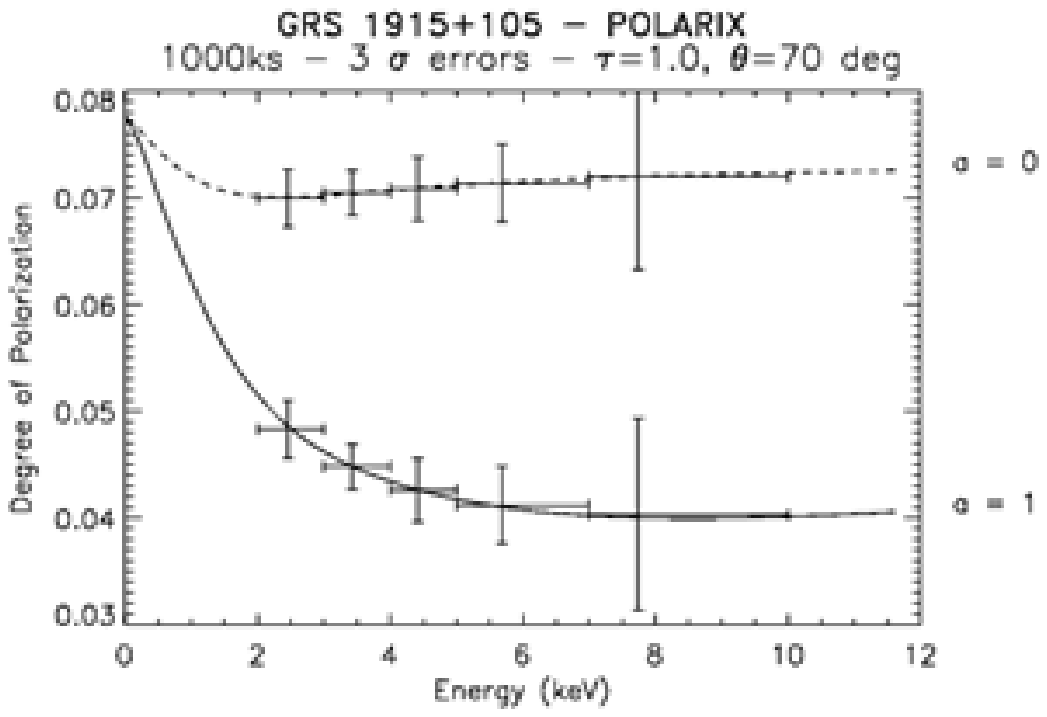}}
\hspace{1cm}
\subfigure[\label{fig:11b}]{\includegraphics[scale=0.5]{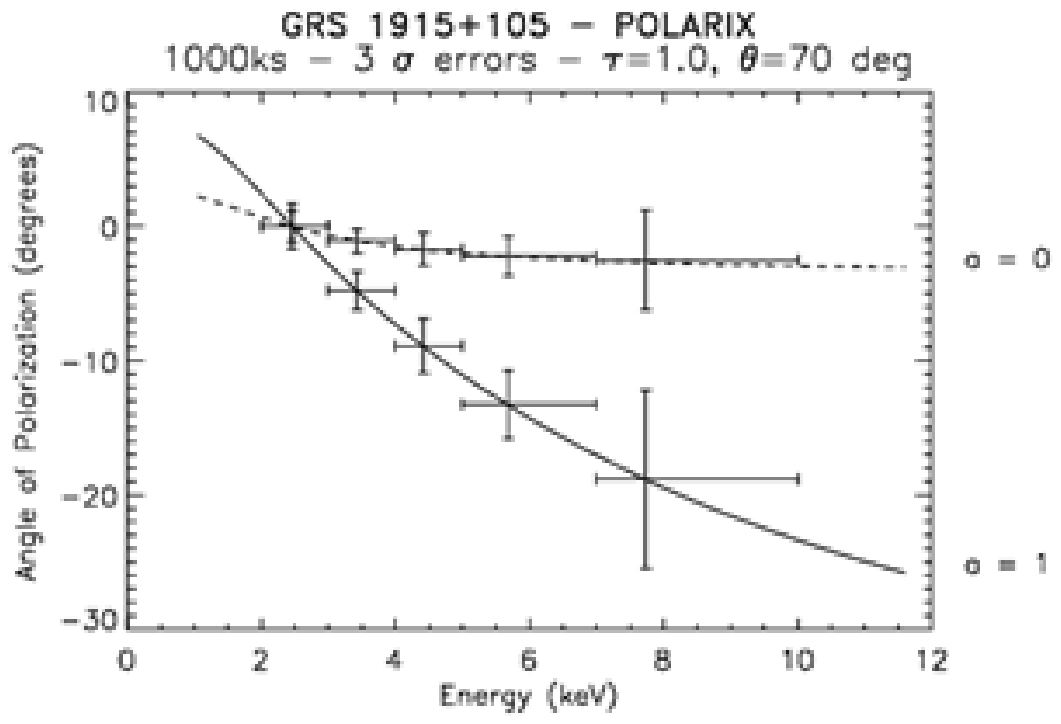}}
\caption{({\bf a}). Polarization degree as a function of energy, expected
 to be measured by POLARIX in GRS 1915+105 (Dovciak et al.2008\cite{Dovciak2008})
The case of a static and a maximally rotating black holes are shown ({\bf b}).
The same but for the polarization angle.}
\end{figure}

\textbf{Galactic Black Hole systems.} When in a high state, the
dominant component in the 2-10 keV emission of Galactic Black Hole
binaries is the thermal emission from the accretion disk. The
innermost regions of the disk are very close to the black hole,
where GR effects are very strong. These effects cause a rotation
of the polarization angle of the radiation emitted from the disk,
the amount of rotation depending on the azimuthal angle and the
radius of the emitting point.  Even after averaging over the
azimuthal angle, a net rotation remains. The  closer the black
hole is to the emitting point, the larger the rotation. Because
the emission is locally a thermal one, and because the temperature
decreases with the disc radius, what is eventually observed  is a
rotation of the polarization angle with energy (Stark and Connors
1977\cite{Stark1977}, Connors et al. 1980\cite{Connors1980},
Dovciak et al. 2008 \cite{Dovciak2008}, Li et al. 2009
\cite{Li2009} Schnittman et al. 2009 \cite{Schnittman2009}) . The
effect is particularly strong for a spinning black hole, where the
disc can extend down very close to the black hole horizon
(fig.\ref{fig:11a}, \ref{fig:11b}). It will be a powerful tool to
study the behavior of radiation in the extreme gravitational field
of the black holes, and so to test the Gravitational Relativity in
the strong field regime. Moreover, it will provide a method to
measure the spin of the black hole.

\begin{figure}
\centering
\includegraphics[scale=0.5] {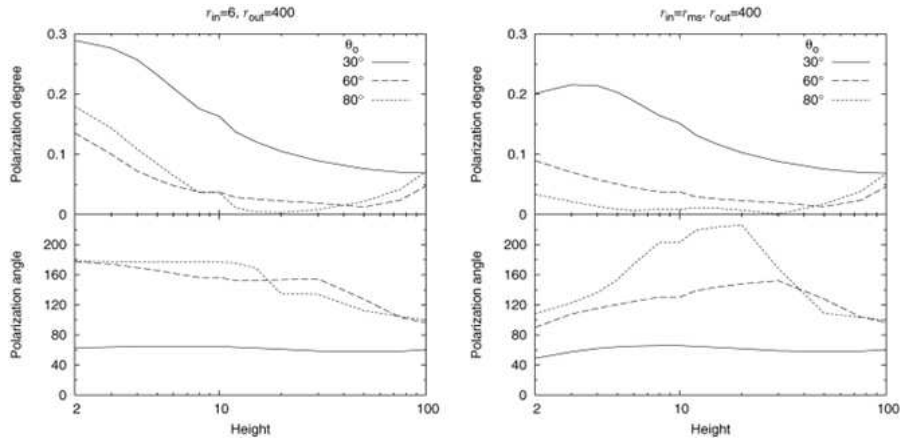}
\caption{The polarization degree and angle as a function of the height
h of the emitting source in the framework of the Miniutti $\&$ Fabian model \cite{Miniutti2004},
for a static black hole (left panel) and a maximally rotating black hole (right panel)}
\label{fig:12}       
\end{figure}

\textbf{AGN.} In AGN, the disk thermal emission is outside the
working band of POLARIX. However, GR effects may manifest
themselves through time-dependent, rather than energy-dependent,
rotation of the polarization angle. In fact, to explain the
puzzling time behavior of the iron line in the famous Seyfert 1
MCG-6-30-15 (the best and most studied case so far for a
relativistic iron line, Fabian et al. 2000 \cite{Fabian2000} and
references therein), it has been proposed (Miniutti $\&$ Fabian
2004\cite{Miniutti2004}) that the primary emission originates in a
small region close to the black hole spinning axis, the observed
variability being due to the variation of the height of the source
(fig.\ref{fig:12}). If this is indeed the case, the polarization
degree and angle of the reflected radiation must also vary in a
characteristic way (Dovciak et al. 2004\cite{Dovciak2004}), which
depends on the spin of the black hole. Polarimetry can therefore
provide a further and powerful probe of radiative transfer in a
strong gravity field as well as an estimate of the black hole
spin. In MCG-6-3-15, a MDP of about 4$\%$ can be reached with
POLARIX in 300 ks. A long look (1 MS or more) to this source may
provide a first test of the model.

\subsubsection{Quantum Gravity}
One of the most ambitious efforts of modern physics is to develop a theory that unifies
Gravity with the other three forces within a single theoretical framework.
Different approaches to Quantum Gravity are pursued (Loop, String, non commutative
space-times) all sharing the general problem of finding good observational tests
 (Amelino-Camelia 2004\cite{Amelino2004}).

One of such few tests can be done with polarimetry. Loop Quantum
Gravity predicts that, at the Planck scale, a small birefringence
effect is present which, for linearly polarized radiation, results
in a rotation of the polarization angle along the photon path
(Gambini $\&$ Pullin, 1999\cite{Gambini1999}). The rotation of
the polarization angle is proportional to the distance of the
source, and to the square of the energy of the photon, via an
adimensional factor of proportionality, $\eta$. Previous UV and
X-ray polarization measurements (the latter performed on the Crab
Nebula, the only source for which there is a positive detection so
far) have already constrained  $\eta$ to be less than about
10$^{-4}$ (Gleiser $\&$ Kozameh 2001\cite{Gleiser2001}; Kaaret
2004 \cite{Kaaret2004}) with claims as low as a few times
10$^{-7}$ based on optical/UV measurements of a Gamma-ray burst
afterglow (Fan et al., 2007)\cite{Fan2007}. Even tighter
constraints to $\eta$ of order 9 $\times$10$^{-10}$ have recently
been put by Maccione et al. (2008)\cite{Maccione2008} comparing
the recent polarization measurement of the Crab pulsar in soft
gamma-rays with INTEGRAL with the optical observations (6 $\times$
10$^{-9}$ if only the gamma-ray result is used.)

A deeper test of the theory would require, on the one hand,
measuring (or putting upper limit to) values of $\eta$ as small as
possible; on the other hand, to check results against e.g.
possible errors due to intrinsic polarization angle variability by
looking to different sources at different distances.   Several
bright enough Blazars at different distances are available to put
the result on a firm statistical basis.

\section{POLARIX design}
\subsection{Mission Description and Design Drivers}

POLARIX is composed of four basic elements: (1) the Service Module
(2) three (possibly five) telescopes with 3.5 m focal length, (3) the Mirror Modules
which accomodates the telescopes, inserted into the
Service Module, (4) the Focal Plane Array composed
by a structure containing the Gas Pixel Detectors (GPD).
POLARIX will be able to detect in 10$^{5} s$ a polarization higher than
10 $\%$ for a 1 mCrab source, with an energy resolution of 20$\%$ at 6 keV, an angular
resolution of about 20 arcsec and a field of view of 15 arcmin x 15 arcmin.  In the following
paragraphs we will describe in detail the components of POLARIX.

\subsection{Philosophy}
POLARIX capabilities are determined by the performances of the
telescopes combined with the GPD. Both the
telescopes and the GPD are already existing and qualified.

The general philosophy driving the design of POLARIX is to have a
mission fully exploiting these capabilities, while remaining
within the (ambitious) budget limitations fixed by the Italian
Space Agency (ASI) Announcement of Opportunity (AO) for Small
Missions.  We coped with this tight economic constraint thanks to :
\begin{itemize}
\item a satellite platform as standard as possible  \item
commonalities with other missions   \item a plug and play philosophy of
Assembly Integration and Verification (AIV) (to simplify  Payload/ Service Module  integration). \item a
Ground Segment organization assigning a major role to science
institutes.\item a minimal mission duration \item a prime contractor
role assigned  to a scientific institution (INAF).
\end {itemize}

\subsection{Optics}
As part of the JET-X project, four mirror modules were
built, 3 flight units (FM) and an Engineering Qualification
Model (EQM). The EQM has been
used for the qualification test campaign, but it has the same
characteristics of the 3 FM units. One of these units is shown
in fig. (\ref{fig:13}) below. These mirror modules were
developed at the Brera Observatory and were manufactured by
Medialario. They have 12 concentric gold-coated electroformed
Ni shells with a focal length of 3.5 m. The shells are 600 mm
long with diameters ranging from 191 to 300 mm. The effective
area and point spread function of these mirrors have been
measured at the Panter facility for a range of
energies and off-axis angles. The last calibration campaigns
were performed on the third flight model, FM3, that is
now onboard the Swift satellite. As a standalone unit they
were measured the last time at the Panter facility in July
2000. A calibration image of two sources displaced by 20
arcseconds (Fig.\ref{fig:14}) immediately shows the image
quality of these mirrors and their capability to separate two
nearby sources. The total effective area of a single unit is
$\sim$ 159 $cm^{2}$ at 1.5 keV and $\sim$ 70 $cm^{2}$ at 8 keV, while the
Half Energy Width (HEW) is $\sim$ 15 arcsec at 1.5 keV and
$\sim$ 19 arcsec at 8 keV (Citterio et al. 1996  \cite{Citterio1996}).
For the most recent results on the inflight calibrations of
the mirror unit now flying onboard Swift see Moretti et al.
(\cite{Moretti2005}) and Romano et al. (\cite{Romano2005}).
The total mass of each unit is
of 59.9 $\pm$ 0.5 kg. The maximum diameter of one Mirror unit
occurs at the interface flange and it corresponds to a diameter of
388 $\pm$ 4 mm. The maximum height is 667.5 $\pm$ 5.0 mm. The
interface mounting of each unit has a planarity of $<$ 0.02 mm
and a circular shape ring of 20 mm diameter. Each unit
consists of a forged stainless steel cylinder, fitted at each
end with a stainless steel shell support or spider.
The mounting interface flange is at the center section of the
cylinder. The twelve electro-formed nickel mirror shells are
mounted concentrically within the mirror cylinder, controlled
by grooves machined into the spider units. The mirror
alignment reference flat is carried at the center of the
forward spider. Both spiders also carry the required beam
stops and blanking plates. The thermal environment for the
Mirror units must be such that the HEW would not be degraded by more
than 10 arcsec. This implies that the mirror units should have
a thermal gradient of less than 2$^\circ$C.

\begin{figure}
\centering
\includegraphics[scale=0.5] {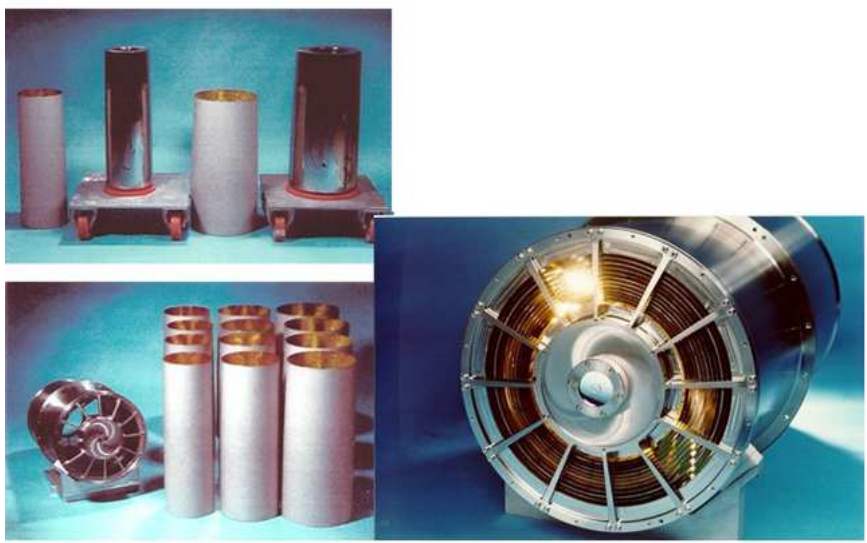}
\caption{One of the four mirror modules built for the
JET-X telescope (on the right). It consists of 12 nested
 Wolter-I grazing incidence mirrors held in place by
 front and rear spiders (see bottom panel on the left side).
 In the left top panel there are shown also two mandrels
 used for the electroformation of the shells, together with two shells.}
\label{fig:13}       
\end{figure}

\begin{figure}
\centering
\includegraphics[scale=0.5] {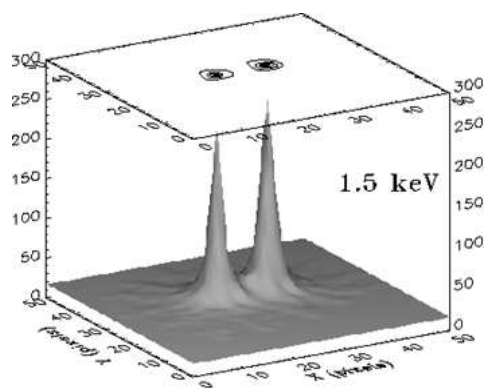}
\caption{Image of two point sources displaced by 20 arcseconds,
 made during mirror calibration at the Panter calibration facility. The image is taken at 1.5 keV.}
\label{fig:14}       
\end{figure}

\subsection{Focal Plane}
The Focal Plane (FP) structures are composed of a detector mounting
plane and a sunshield; the interface with the lightshield tube
(described below) is provided by a continuous Al ring.  The lightshield tube
encircles all the detectors and connects the FP to the service
module and to the optics module. The harness (power, data) will
run from the Focal Plane to the Service Modulus through the
lightshield tube.

\begin{figure}[htpb]
\centering
\subfigure[\label{fig:18a}]{\includegraphics[scale=0.2]{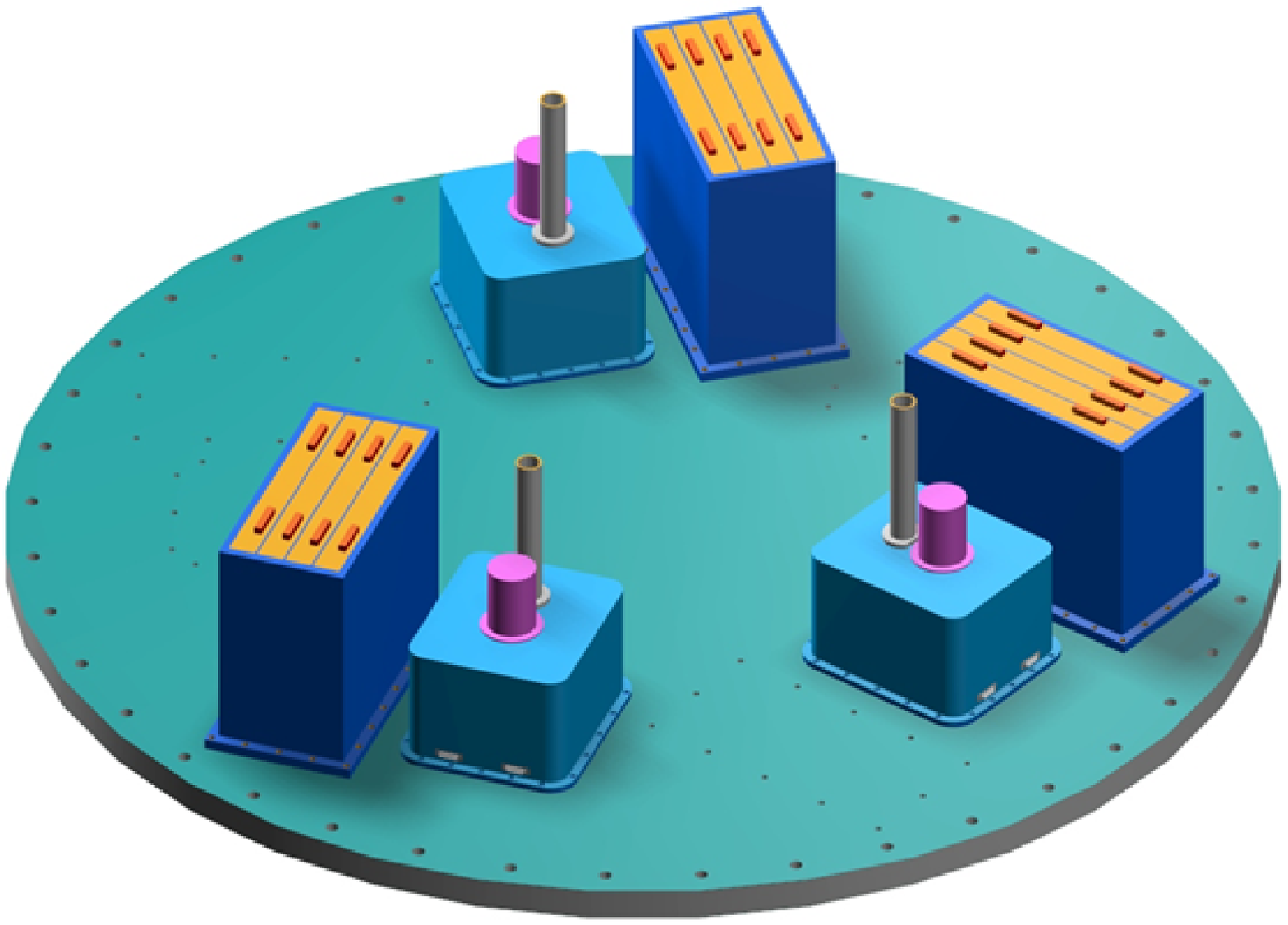}}
\hspace{1cm}
\subfigure[\label{fig:21b}]{\includegraphics[scale=0.2]{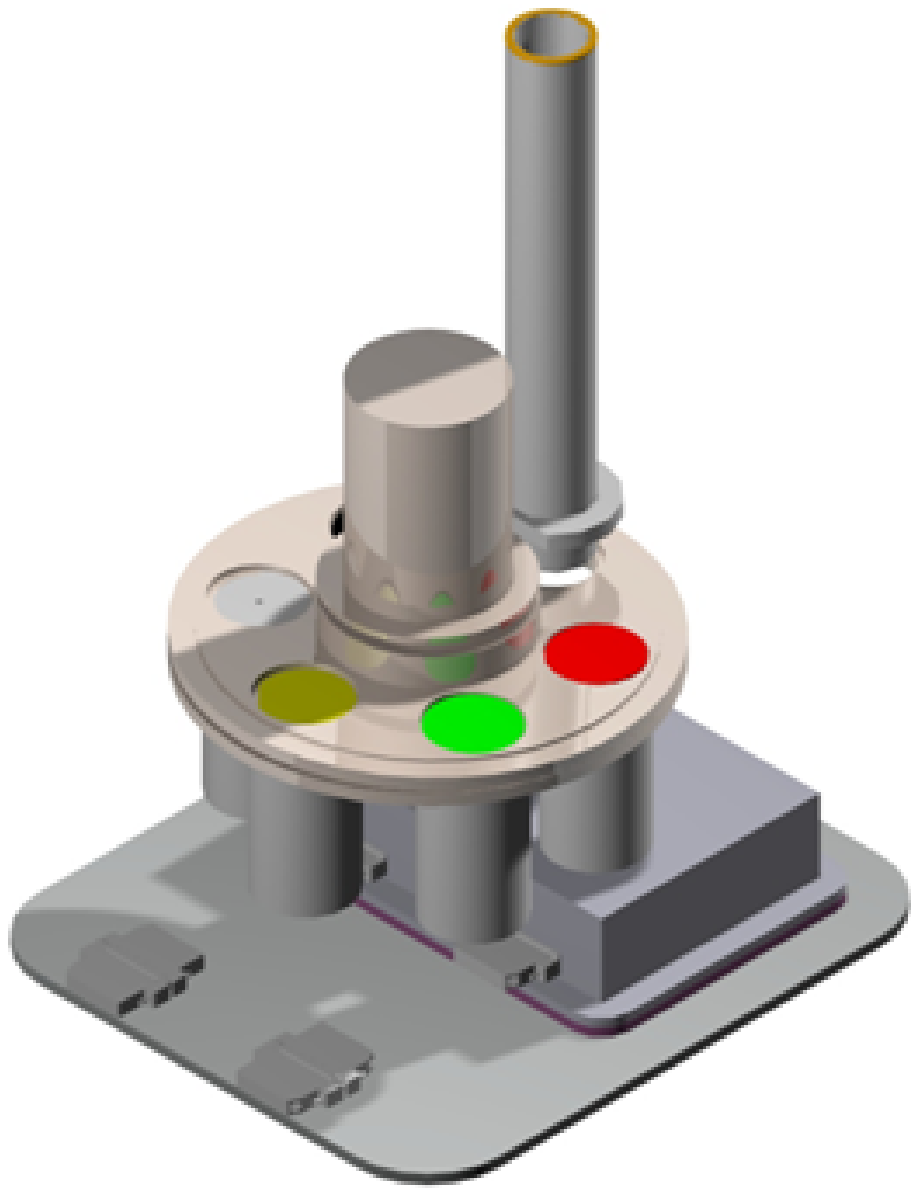}}
\caption{({\bf a}). Focal plane layout  ({\bf b}). Front End layout}
\end{figure}

The Focal Plane contains three Front Ends and the associated
Back End electronics (BEE, Fig.\ref{fig:18a}). Each Front End unit
contains the detector, a filter wheel with calibration sources
(see next paragraphs) and a baffle with an electrostatic grid
on top.

The BE electronics units, one for each detector unit,  are
close ($<$ 20 cm)  to the detectors, (fig.{\ref{fig:21b}). The
High Voltage Power Supply (HVPS) are inside the BEE boxes to reduce the noise.

Each detector is inserted in a dedicated box which will be
thermally controlled by a Peltier at a mean temperature
in the 10$^\circ$-20$^\circ$C range with a max
oscillation of $\pm$ 1$^\circ$C. The weight of the focal plane
(excluding the harness) is shown in table \ref{tab:1}.

\begin{table}
\caption{Weight of the focal plane components}
\label{tab:1}       
\begin{tabular}{llll}
\hline\noalign{\smallskip}
Unit name & Quantity & Mass (kg) & Total (kg) \\
\noalign{\smallskip}\hline\noalign{\smallskip}
Detector & 3 & 0.4 &  \\
FW,baffle, box & 3 & 2.85 &   \\
Front-end total & 3 & 3.25 &  \\
Back-end electronics & 3 & 3.5 &  \\
Focal plane tray & 1 & 4.5 &  \\
Focal plane sun-shield & 1 & 9 & \\
Total     &   &   & 33.55\\

\noalign{\smallskip}\hline
\end{tabular}
\end{table}

\subsubsection{Detectors}
The purpose of each focal plane instrument  is to provide, in
the energy range 2-10 keV, polarization measurements
simultaneously with angular, spectral and timing (at few
$\mu$s level) measurements. Each instrument  is based on a Gas
Pixel Detector\cite{Bellazzini2006}\cite{Bellazzini2007}, a
position-sensitive counter with proportional multiplication
and a finely subdivision of the charge collecting electrode in
such a way that photoelectron tracks can be accurately
reconstructed, as shown in fig. \ref{fig:track}.

\begin{figure}
\centering
\includegraphics[scale=0.5] {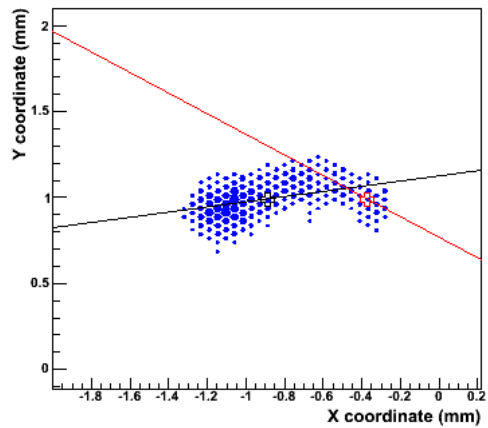}
\caption{Example of real track collected by the polarimeter}
\label{fig:track}       
\end{figure}

The Gas Pixel Detector is an advanced
evolution of the MicroPattern Gas Chamber. It is based on a
gas cell with a thin entrance window, a drift gap, a charge
amplification stage and a multi-anode readout which is the
pixellated top metal layer of a CMOS ASIC analog chip (105600
pixels at 50 $\mu$m pitch). The sealed GPD is shown in
fig.\ref{fig:15a}. The different components are shown in
fig.\ref{fig:15b}. When the X-ray photon is absorbed in the
gas gap the ejected photoelectron produces an ionization
pattern in the gas (track). The track is drifted by a uniform
electric field to the Gas Electron Multiplier (GEM) where the
charge is amplified. The linear polarization is
determined from the angular
distribution of the photoelectrons as derived from the
analysis of the tracks. The analysis algorithm reconstruct the
impact point with a precision of $\sim$ 150 $\mu$m FWHM,
largely oversampling the PSF. The FOV is 15 $\times$ 15
square arc minutes. The effect which mostly affects the
resolving power is the blurring due to the transversal diffusion
in the gas of the ionization track along the drift path to the
collecting electrode.

Below the GEM, at a distance less than a few
hundred micron, the top layer of the multilayer ASIC is
covered with metal pads with a high filling factor distributed
on a hexagonal pattern. Each pad is connected underneath to
its own independent analog electronic channel. The average
noise of the electronics is only 50 $e^{-}$ rms. With a
moderate gain of 500, single electrons produced in the gas
cell can be detected. The system has self-triggering
capability and only the charge of the pads included in a
window around the pixels which have been triggered are read-out and
digitally converted.  In this way  data volume and read-out
time are significantly reduced, and  only sub-frames of 400 to
600 pixels (the so called Region of Interest, ROI), which
include the track completely, are extracted in real time at
each event.

\begin{figure}[htpb]
\centering
\subfigure[\label{fig:15a}]{\includegraphics[scale=0.5]{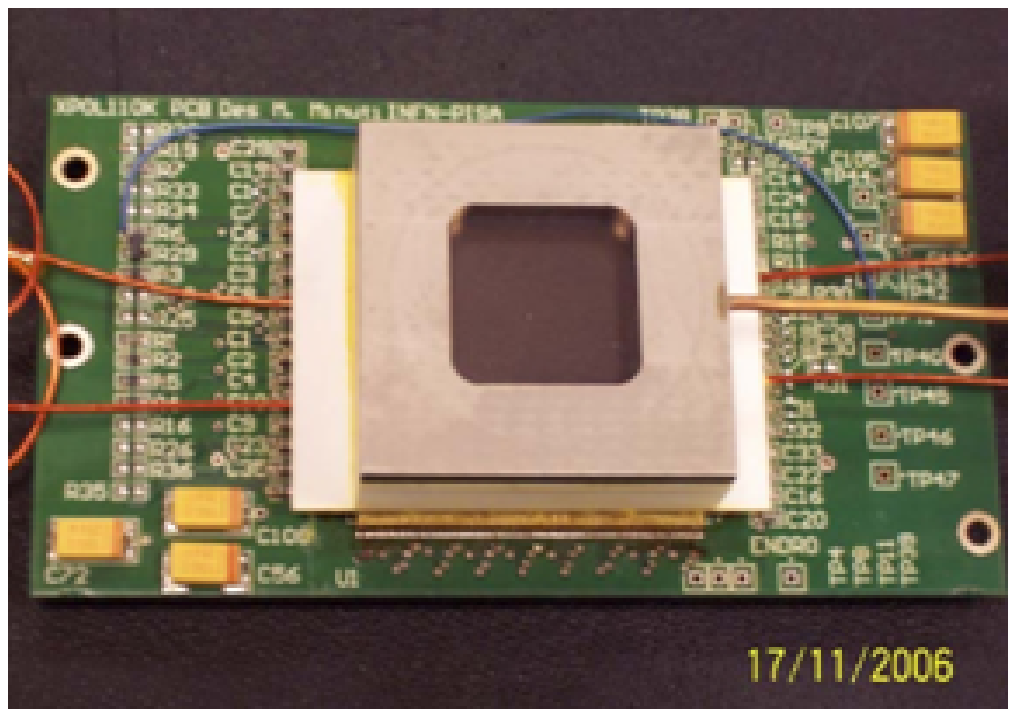}}
\hspace{1cm}
\subfigure[\label{fig:15b}]{\includegraphics[scale=0.5]{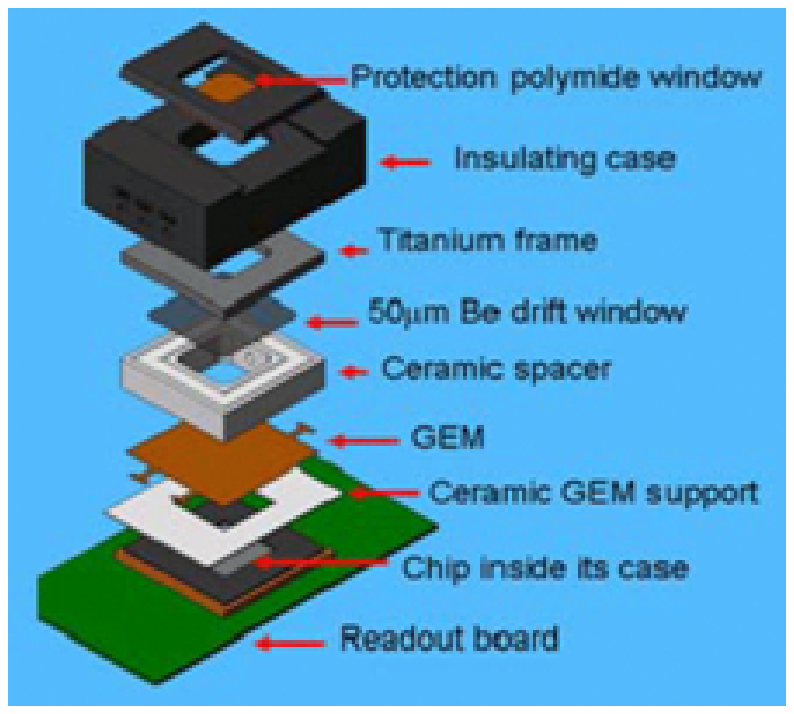}}
\caption{({\bf a}). The sealed gas pixel detector currently working in
laboratory. ({\bf b}). An exploded view of the gas pixel detector.
The window is made of beryllium with an exposed area of 1.5 cm x 1.5 cm and
a thickness of 50 $\mu$m}
\end{figure}

From the point of view of construction, the GPD is a conventional
proportional counter coupled to a VLSI chip. From the existing data
on ageing tests made with mixtures based on noble gases and various
quenching, including Dimethyl Ether (DME), the GPD gas mixtures can withstand the
radiation levels for POLARIX. The major problem can be
the long term pollution of the mixture itself. However, since a long
experience exists on sealed gas counters operating in space for
years, there is strong confidence that, with the procurement of
very pure gas and a proper selection of materials to be used
inside the detector, the problem can be overcome. Long term
stability in a sealed gas cell with Beryllium window was achieved
not only for proportional counters (e.g. COS-B, HEAO-1, Einstein,
GINGA, XTE) but also for Gas Scintillation Proportional Counters
(GSPC, TENMA, ASCA, SAX) which are
extremely sensitive to pollution from out-gassing. Presently, a
sealed detector is working for more than two years without showing
any degradation of the performances. All the construction
procedures are compatible for use in space. The GEM is a
metal-coated Kapton film, similar to that used in the Stellar
X-ray Polarimeter (SXRP) detectors.
The detector is a sealed body with no gas refilling system. It is
self-standing with a total weight of a few hundred grams, mainly
due to the HV distribution, potting and connectors. The gas volume
(10 mm absorption/drift gap), as the baseline, is filled with
20$\%$He-80$\%$DME at 1 bar. Based on studies in progress, the
pressure could be increased to 2 bar to improve the efficiency
especially at higher energies. The mixture could also be
substituted with other mixtures, less sensitive to lateral
diffusion during drift, to allow for thicker absorption/drift gap.
In this case the overall length could increase by 10 mm. The
baseline window is 50 $\mu$m of Beryllium. If technology
will show that an improved performance can be achieved at energies below
1.5 keV, a thin plastic window could be adopted. The read-out VLSI
ASIC chip, based on 0.18 $\mu$m CMOS technology, has been already
successfully tested. The self-trigger download mode works
perfectly. All the major functions have been already tested and
are compliant with the requirements for the GPD.

\subsubsection{Radiation hardness}
A major problem found in past X-ray missions with gas
detectors with multiplication was the effect of highly charged
particles on the electrodes. The passage of heavy nuclei in
the multiplication regions may give rise to self-sustained
sparks that can damage or even destroy the electrodes. This
was a serious problem for, e.g., XTE/ASM or INTEGRAL/JEM-X,
resulting in a reduced performance of the detectors;
in EXOSAT/PSDs it was the likely cause of the failure of
both gas detectors. This problem is usually prevented by introducing
limitations to the current that can flow in the stage. In any
case, the criticality is there whenever the detector is
operated to a gain level too close to the break-down level.
Thanks to the low noise and to the fully pixel concept we
operate the GEM at a very low gain, orders of
magnitude from the break-down value. In any case, we extensively
tested a GPD with X-ray fluxes comparable to three years of
operation in space. The most significant test was
performed at Heavy Ion Medical Accelerator in Chiba in Japan.
The GPD was irradiated with Protons (E $<$ 160 MeV), He, C,
N,O, Ne, Si, Ar, Fe and Xe. Tests with Fe were particular
significant. GPD survived to a dose of Fe ions of 500 MeV/n,
equivalent to that expected in 40 years of operation in a low
Earth orbit.

\subsubsection{Back-End electronics}
The ASIC output consists of the information about ROI
coordinates and the charge collected by each pixel within the
ROI. Coordinates are expressed by four binary words
(xmax[8..0], xmin[8..0], ymax[8..0], ymin[8..0]), while charge
is given as Front End analog output. At the trigger a ROI is
internally indexed and the control electronics can begin to
read it out.  The charge of each pixel is serialized on a
common analog output. A differential output buffer is used to
drive this signal to the ADC.

\begin{figure}
\centering
\includegraphics[scale=0.6] {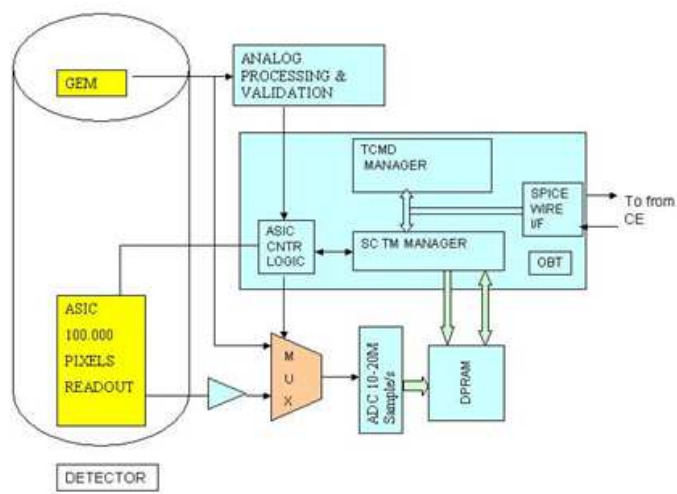}
\caption{Diagram of the interface BE  electronics}
\label{fig:16}       
\end{figure}

The block diagram of the BEE is shown in fig.\ref{fig:16}. The
analog output from ASIC will be A/D converted at the interface
BEE. The ASIC will be controlled by a dedicated
FPGA, the A/D converted data will be zero-suppressed,  a
microprocessor will manage the telecommand and the
housekeeping (HK). The interface BEE will also assign
the time to each event. The BEE is the
assembly that contains the electronic boards placed between the
detector assembly (Front-End Electronics (FEE) = GEM + Readout ASIC)
and the Control Electronics box (CE).

It is responsible for :
\begin{itemize}
\item Distributing and filtering the power supply required to
    the low voltage front-end electronics;
\item Supplying the detector with all the high voltages needed; \item
Managing the Front-End Electronics; \item Implementing a spectroscopy
electronic chain for the GEM analog output; \item Digitally
converting the analog output of the FEE (ADC function); \item Storing
auxiliary information related to each event (e.g. X,Y coordinates
of the ROI corner); \item Time-tagging the events with 8 $\mu$s of
resolution and 2 $\mu$s of accuracy with respect to the Universal
Time (UT) (from GPS); \item Digitally performing some basic
processing (pedestal calculation, suppression of not-fired pixels)
\item Temporarily storing the converted data (both from ASIC and
GEM); \item Integrating some HK and Science Ratemeters related to
the detectors activity (e.g. good event, rejected events, …);
\item Providing Instrument HK to the CE for active monitoring and
telemetry purposes. \item Implementing the Peltier Driver for the
detector temperature control.
\end{itemize}
For analog signal integrity reasons, this back end assembly is
placed close to the detector assembly.

\subsubsection{Calibration sources} The detectors will be accurately
calibrated on ground. Although available data show a good stability with
time, we want to preserve the capability to calibrate
periodically during the flight in order to check the long term
stability for possible change of the components and/or for the
possible aging of the GEM or of the filling mixture. We want
to calibrate the efficiency, the gain, the energy resolution
and the modulation factor for a minimum of two values of the
energy. The calibration will be done with both unpolarized
source and polarized source (Muleri 2007, \cite{Muleri2007}).
Calibration sources will be mounted in three positions of the
filter wheel.

In the following we describe both types of calibration
sources.
\begin{itemize}

\item Polarized calibration source.

        The modulation factor of the polarimeter
    increases with energy because the tracks
    become longer and straighter, since the specific
    energy loss  and the effect of Rutherford scattering
    decrease. A simultaneous measurement at two energies is
    highly desirable to monitor the behavior of the
    polarimeter in space. Bragg diffraction of X-ray lines
    or continuum tuned at nearly $45^{\circ}$ provides nearly
    $100\%$ polarized radiation. To generate X-rays in
    space, instead of an X-ray tube, we can use a
    fluorescence source of adequate intensity such as
    $Fe^{55}$. Its Mn K-lines can excite 2.6 keV line from
    a thin PVC film. Graphite crystals at $45^{\circ}$ can
    reflect those 2.6-keV lines polarizing them nearly at
    100$\%$. LiF crystal reflects the $Fe^{55}$ K-alpha
    line photons close to 47.6 degree polarizing them at 88 $\%$. We
    want to exploit both lines with a composite X-ray
    calibration source. A thin (20 $\mu$m) Graphite
    crystal is attached to a LiF thick crystal to polarize
    simultaneously 2.6 keV and 5.9 keV photons. A sheet of 40
    $\mu$m thick PVC crystal is placed in front of the
    $Fe^{55}$ source to convert part of the X-ray photons
    into 2.6 keV Chlorine photons. The unabsorbed
    $Fe^{55}$ X-ray photons will cross the graphite
    crystal to be reflected by the LiF. The source will be
    compact (small volume and weight) to allow for a safe use
    in space. The monitoring of the modulation factor
    during the observation will also allow us to check
    whether any pollution has altered the drift (and
    therefore the diffusion coefficient) in the gas
    mixture.

\item Unpolarized calibration source.

        The gas gain of the detector is a function of the
    voltage difference across the GEM. Pollution of the
    gas due to outgassing and ageing of the gas mixture can
    require higher voltage difference to reach the same
    gas gain. We want to monitor the gas gain using two
    radioactive sources. One radioactive source will be
    Fe$^{55}$ . The Fe$^{55}$ photons impinge into the
    whole detector surface to monitor the gain across it.
    The counting rate will be 20 c/s. $Fe^{55}$ can be a
    point source or a source diffused in a circular
    surface. The second X-ray sources will be  Copper
    (8.04 keV) photons extracted by a radioactive
    Cd$^{109}$ source. This second X-ray source will be
    used to monitor the linearity of the gas gain with
    time. The counting rate of this second source will be
    10 c/s.

\end{itemize}

\subsection{Control Electronics}
The Control Electronics (CE) is a data handling unit dedicated to
the on-board data processing and the power Scientific Instrument
control. The CE constitutes the central node of the Payload,
managing the data and the power interfaces towards the scientific
detectors and the spacecraft. The unit high level architecture is
shown in fig.\ref{fig:17}.

\begin{figure}
\centering
\includegraphics[scale=0.5] {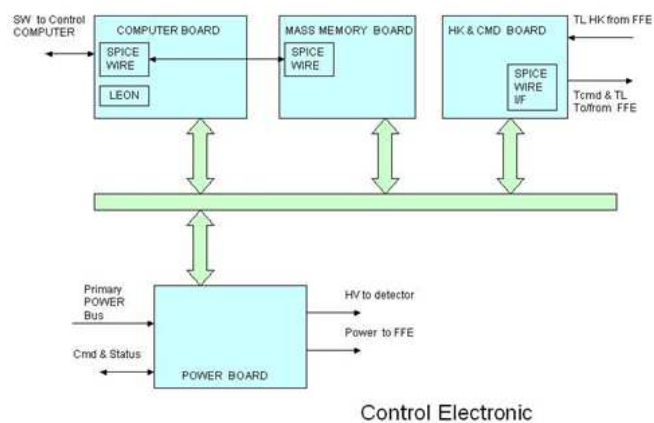}
\caption{Block diagram of the Control Electronics.}
\label{fig:17}       
\end{figure}

The CE is aimed at performing the following principal tasks:

(1) parse and execute the TCs coming from the S/C; (2) generate
scientific and HK telemetries; (3) generate messages, warnings and
errors reports; (4) manage the Payload Instrument Operative Modes
(Boot, Maintenance, Idle, Observation and Test); (5) implement the
detector thermal control algorithms; (6) control the positioning
of three Filter Wheels; (7) A to D convert all the Payload analog
HK lines both for thermal control, position control and HK
purposes; (8) perform the Instrument Control Function, i.e. the
active monitoring of some Payload safety critical parameters in
order to implement a nearly real-time reaction to avoid damages;
(9) in the case of high science data rates, store this data into a
Payload Mass Memory (implemented inside the CE); (10) perform
science data processing.

\subsubsection{Scientific Data Acquisition}
The observation program foresees both faint and bright sources,
with the constraint due to the available average telemetry bandwidth
(50kbit/s).

For the purpose of sustaining high data rates of bright sources,
two strategies are available with the proposed CE architecture: (1)
temporary data storage on a Payload Mass Memory; (2) on board
track reconstruction to transmit on ground only the main photon
parameters.

The Payload Mass Memory will be used only when necessary in
order to bufferize scientific data that exceeds the available
telemetry bandwidth.

The difference between the two strategies is that the mass memory
can only differ the delivery time of the generated data, whilst
the on board track reconstruction is able to reduce the
transferred data for each photon: the only disadvantage is the
loss of some information (raw data are more complete than photon
parameters only).

\subsubsection{Scientific Data Processing}
The on-board track reconstruction is thought to dramatically
reduce the science telemetry volume. In fact in this case, instead
of sending to ground TM packets in which the photon information is
represented by all the pixels in the track (address and
charge; the average number of pixel per photon is 50),
only the photon polarization characteristic (photon
impact coordinates, energy, emission angle and a few quality
parameters allowing for off-line further selection of events) is
transmitted. It
can be estimated that in case of on-board track reconstruction,
the average science TM flow will be reduced by a factor 8. As explained in
the next paragraph, the on-board track reconstruction implies the
use of a DSP. However the baseline for the CE foresees a TSC21020
DSP for two main reasons: (1) it involves a
consolidated architecture (for example it was successfully used on
AGILE CE, MARSIS and SHARAD programs)(2) this architecture is
powerful and has a minimum impact on the hardware if the on-board
processing option is chosen.

The mathematical algorithm at issue implies the estimate of the
first, second and third moment of the track by executing a series of
calculation loops involving floating point operations.

The computational load determined by the execution of the track
reconstruction algorithm has been evaluated for the TSC21020. The
result of the analysis proved that the selected DSP is capable of
sustaining the event peak rate for the POLARIX mission
determining a CPU occupation lower than 30\%.

\subsection{Payload main budgets}
Table \ref{tab:22} summarizes the results of the telemetry
data rate estimate with special focus on the Mass Memory
occupation. In table \ref{tab:19} the POLARIX payload power budget
and mass budget are shown.

\begin{table}
\caption{Telemetry data rates and Mass Memory usage.}
\label{tab:22}       
\begin{tabular}{llll}
\hline\noalign{\smallskip}
 & with zero-suppr. & with on-board analysis & note  \\
 Average time rate  & 35 kbit/s & $\sim$ 4.25 kbit/s & 30 ev/s \\
 Peak time rate  & 231 kbit/s & 28.1 kbit/s & 200 ev/s\\
Mass memory usage after 1day @ ptr  & 100$\%$ & 12.2$\%$ & \\
\noalign{\smallskip} \noalign{\smallskip}\hline
\end{tabular}
\end{table}

\begin{table}
\caption{Power and mass budget. In the back-end electronics the power for mirror
temperature control is excluded, the power for detector thermal
control is excluded. Assumed 2W
of secondary power for each detector front-end board. For PPS
Generator Power the range is 7-11 W, we assumed 10W. The Mass range is 4-6Kg,
we assumed 5Kg. In the total the  harness excluded}
\label{tab:19}       
\begin{tabular}{llllll}
\hline\noalign{\smallskip}
Unit name & Quantity & Primary power (W) & Total Primary Power (W) & Mass (kg) & Total Mass (kg) \\
\noalign{\smallskip}\hline\noalign{\smallskip}
Back-End electronics & 3 & 12.9 & 38.7 & 3.5 & 10.5 \\
CE & 1 & 15.3, & 15.3 & 4 & 4 \\
PPS Generator & 1 & 10 & 10 & 5 & 5 \\
TOTAL &   &  -, & 64.1  & - & 19.5\\
\noalign{\smallskip}\hline
\end{tabular}
\end{table}

\section{Mission analysis}
\subsection{Orbit}
The orbit selected for POLARIX is an equatorial circular  LEO (Low
Earth Orbit), achievable by   Vega Launcher (the ESA smallest
launcher). The specific orbital features are the following:

\begin{itemize}
\item Altitude            505$\pm$ 15 km \item Inclination
5$^\circ$ $\pm$ 0.15$^\circ$ \item Sun eclipse duration 36 minutes
max
\end{itemize}

This specific orbit allows the utilization of ASI Ground Station
at Malindi (Kenya). The  POLARIX visibility from the Malindi
Ground station is  8 - 11 min per orbit.

The nominal mission duration is 1 year, plus 3 months for
commissioning and 1 month for decommissioning.  For nominal
operation the satellite can  be supported  by  the Malindi ground
station only. The launch has been considered in  2014 ,   when
solar activity  decreases.

The  selected orbit  features and  the estimated delta V reduction
from atmosphere drag  (mission life-time during  solar minimum
phase)  can be maintained without  any orbital  correction.
Consequently  the satellite does not strictly need a propulsion
sub-system.

\subsection{Launcher}
The launch site is Kourou, latitude 5.0647$^\circ$ and longitude
307.3602$^\circ$. The typical Vega mission includes a three stage
sub-orbital ascent (Vega User Manual, Issue 3/ Revision 0, March
2006).

\begin{table}[!hhh]
\caption{Injection accuracy at $\pm$ 1$\sigma$ for a
Circular Orbit Mission, reference altitude km 700}
\label{tab:18}       
\centering
\begin{tabular}{ll} \hline\noalign{\smallskip}
 \underline{Parameter} & \underline{1$\sigma$ accuracy} \\
\noalign{\smallskip}\hline\noalign{\smallskip}
Altitude & $\pm$ 15 km\\
Inclination & $\pm$ 0.15$^\circ$ \\
Launch time & 3 sec \\
\noalign{\smallskip}\hline
\end{tabular}
\end{table}

The Vega Launch Vehicle can be launched any day of the year and
any time of the day. Performance data for circular orbit missions
with different inclination and altitudes are presented in
fig.\ref{fig.22}. Also table \ref{tab:18} provides the injection
accuracy.

\begin{figure}[!hhh]
\centering
\includegraphics[scale=0.5] {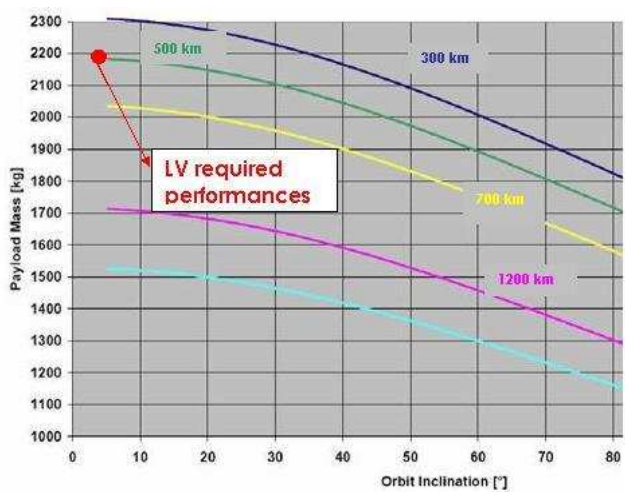}
\caption{Properties of Vega launcher}
\label{fig.22}       
\end{figure}

\subsection{Ground contact and data retrieval}
Ground Contact

The Malindi ground station is optimally located for the near
equatorial POLARIX Satellite orbit. The coverage pattern for this
LEO altitude is a regular sequence of contacts, it is planned to
have 15 contacts a day with an average duration of 8 minutes each and
an average contact period of 120.5 minutes a day. Assuming this
contact time and telemetry high data rate as 512 kbps (with Reed-Solomon
coding), the data volume downloadable is 3.25 Gb/day.

\section{Spacecraft description}
\subsection{Satellite design overview}
Thanks to an intensive  interaction  between the institutes and the
industries a POLARIX mission scenario, compatible with the   Small
Mission  ASI  conditions, has been identified. The  severe cost
limits imposes a "design-to-cost" approach. This approach is
feasible due to the very good  level of  commonalities with  other
projects (Cosmo-SkyMed, Radarsat, Sentinel and GOCE). Moreover
this approach implies the utilization of standard technology and
the selection of the lowest cost option for the system equipments.

In synthesis the  POLARIX  satellite  is composed of the following
main elements:

- the Service Module, which provides all the necessary functions
for the performance of the scientific payload (on-board computer, thermal
control, power generation etc.).

- the 3.5 m  Telescope Structure, which realizes the distance
between the Focal Plane Array  and the Mirrors Modules along their
main axis and support  a  Solar Array  panel (component of Service
Module)

- the Mirror Modules Assembly, accommodate
a mechanical structure  inside the cylindrical thrust  of
Service Module. Each mirror module is equipped with baffle (for
thermal control) and a thin thermal  (for thermal and
contamination control).

- the Focal Plane Array, composed by a structure accommodating
the 3 detectors,  the  Electronics  Interface Unit
containing the functions specific to the detector module and a Sun
shield. Also the payload data handling unit is a component of
Focal Plane Array but,  in order to optimize the mass
distribution,  it is  host in the Service Module.

The Mirror Modules telescopes and the Focal Plane Array make up
the POLARIX scientific payload.

Strong hardware modularity for main elements is considered
with noteworthy advantages on the development schedule. In
particular the payload   can  be developed in parallel with
the other satellite elements.  Final system satellite
integration considers a direct, 'plug-and play' of the main
elements. Fig. \ref{fig:20} shows the POLARIX launch
configuration inside the Vega firing. Fig.\ref{fig:Primo1} is
an exploded view of POLARIX, where  satellite main elements are
identified while fig.\ref{fig:21} is a pictorial view of
POLARIX Satellite.

\begin{figure}
\centering
\includegraphics[scale=0.5] {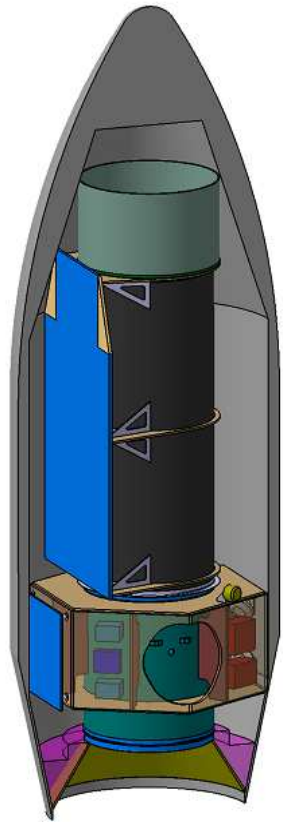}
\caption{POLARIX in the Vega firing}
\label{fig:20}       
\end{figure}

\begin{figure}
\centering
\includegraphics[scale=0.3] {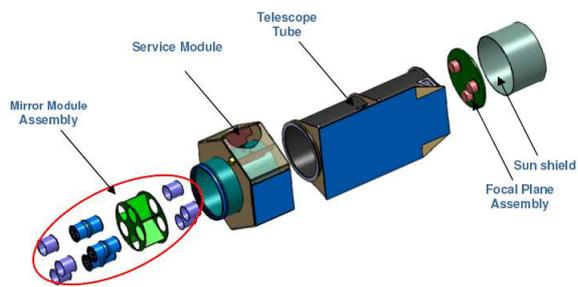}
\caption{Exploded view of  POLARIX}
\label{fig:Primo1}       
\end{figure}

\begin{figure}
\centering
\includegraphics[scale=0.9] {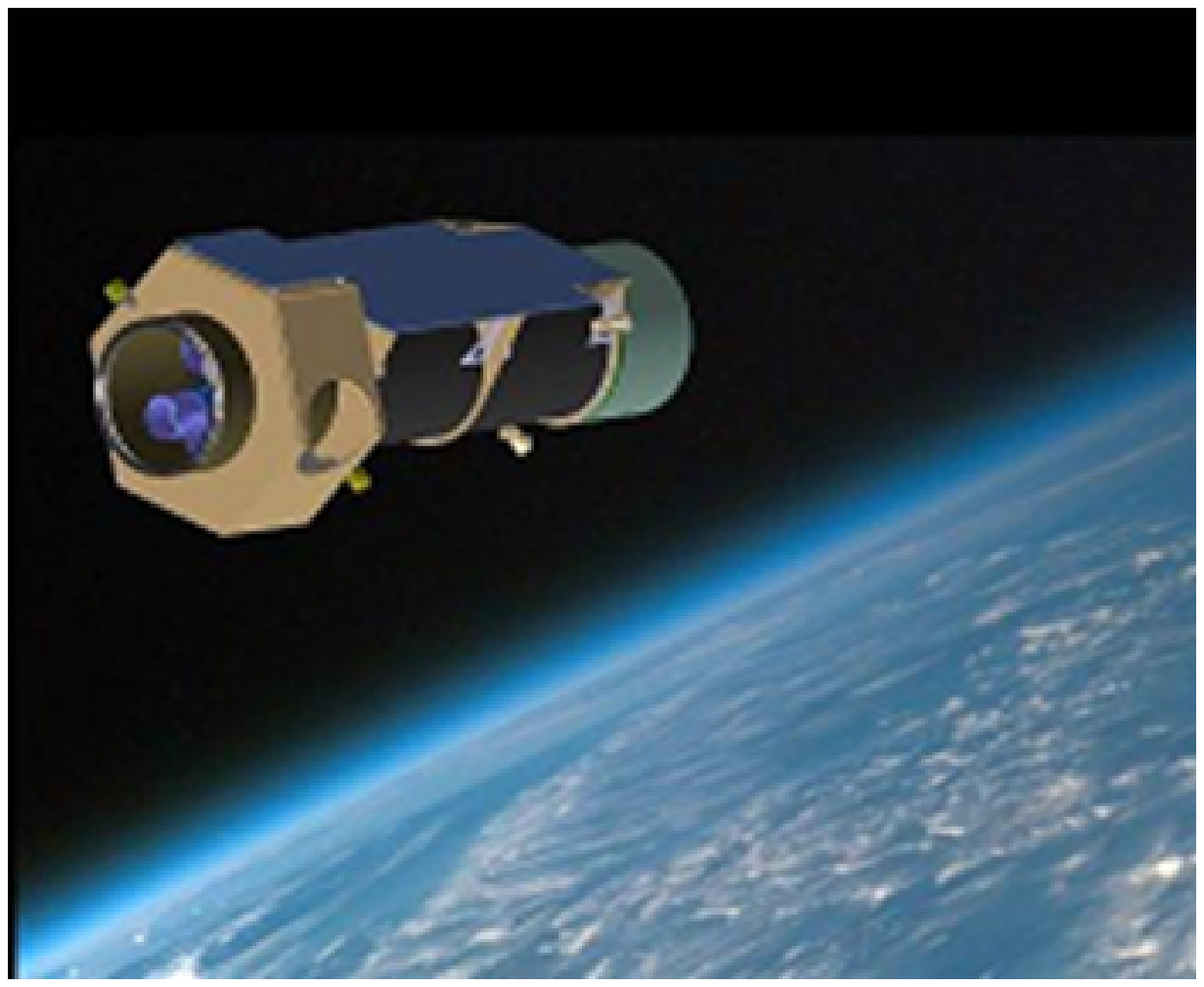}
\caption{Pictorial view of POLARIX}
\label{fig:21}       
\end{figure}

\subsection{Mechanical and thermal design analysis}
Preliminary Mechanical and thermal analysis has been performed to
support the satellite system design. Figures  \ref{fig:30} and
\ref{fig.31} show the Finite Element Model (FEM). On the basis of
FEM the mechanical structure has been designed and the satellite
structure mass consolidated. The mechanical environment simulated
by the satellite's  FEM is compatible with the mechanical design
references of  Mirror and Detector modules.

Preliminary thermal control design has been supported by a
dedicated satellite Thermal Mathematical Model (TMM). The thermal
control items (heaters, thermal blanket, Sun shield)
have been consolidated thanks to various environmental thermal
condition simulated with the TMM.

For  the  Payload thermal control design  the reference payload
temperature are: - Mirror Module : 20 $^{o}$C $\pm$ 2 $^{o}$C -
Detector Module : 15 $^{o}$C $\pm$ 2 $^{o}$C.

The preliminary thermal analysis has demonstrated the feasibility
of payload thermal control by means of a standard approach.
Moreover  the max heater power request estimated by the
preliminary thermal model is the following (three telescope
configuration case) :

- Mirror Assembly :     45 W - Detector Array :      73 W -
Service module :  50 W

\begin{figure}[htpb]
\centering
\subfigure[\label{fig:30}]{\includegraphics[scale=0.5]{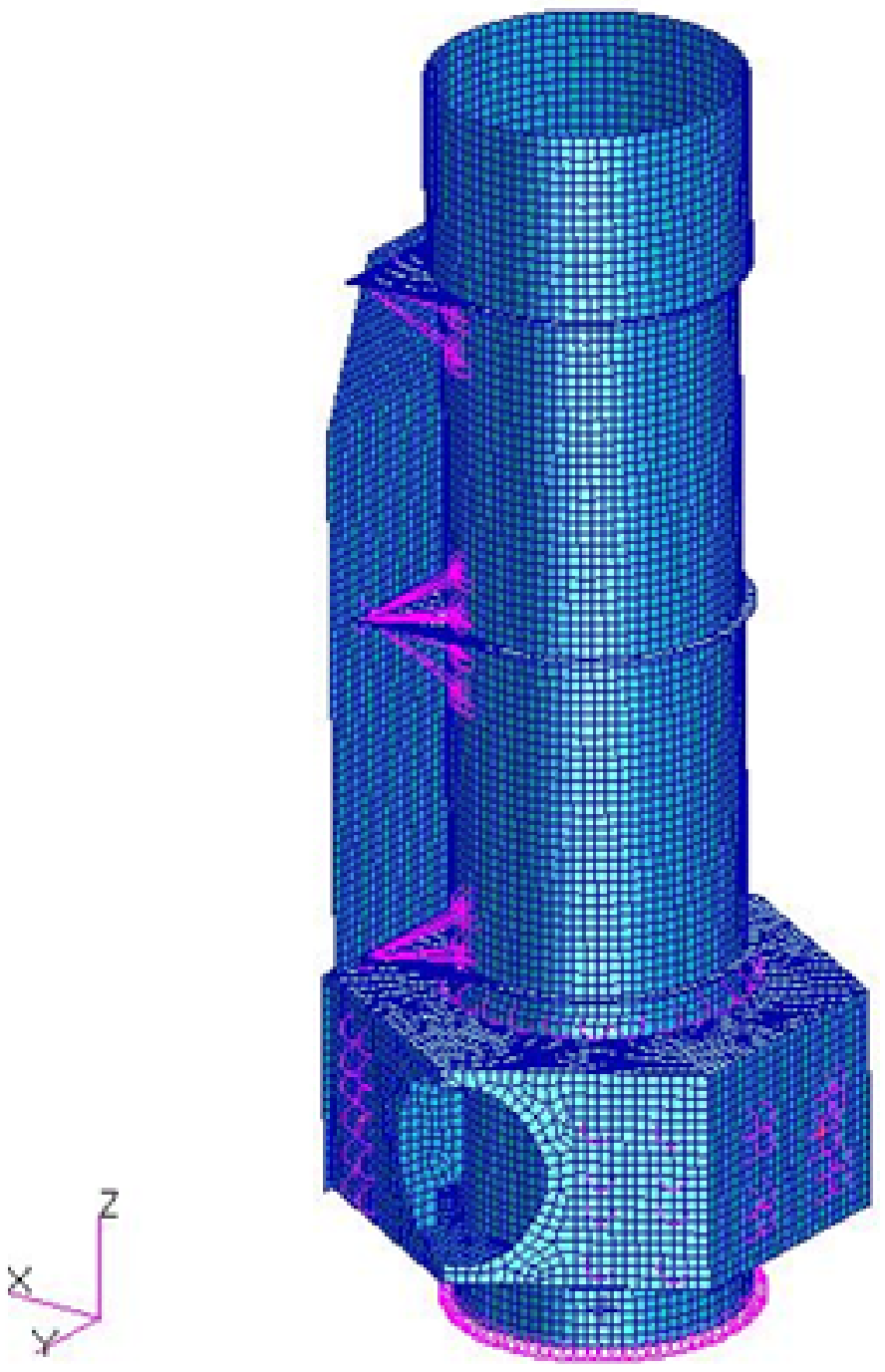}}
\hspace{0.3cm}
\subfigure[\label{fig.31}]{\includegraphics[scale=0.5]{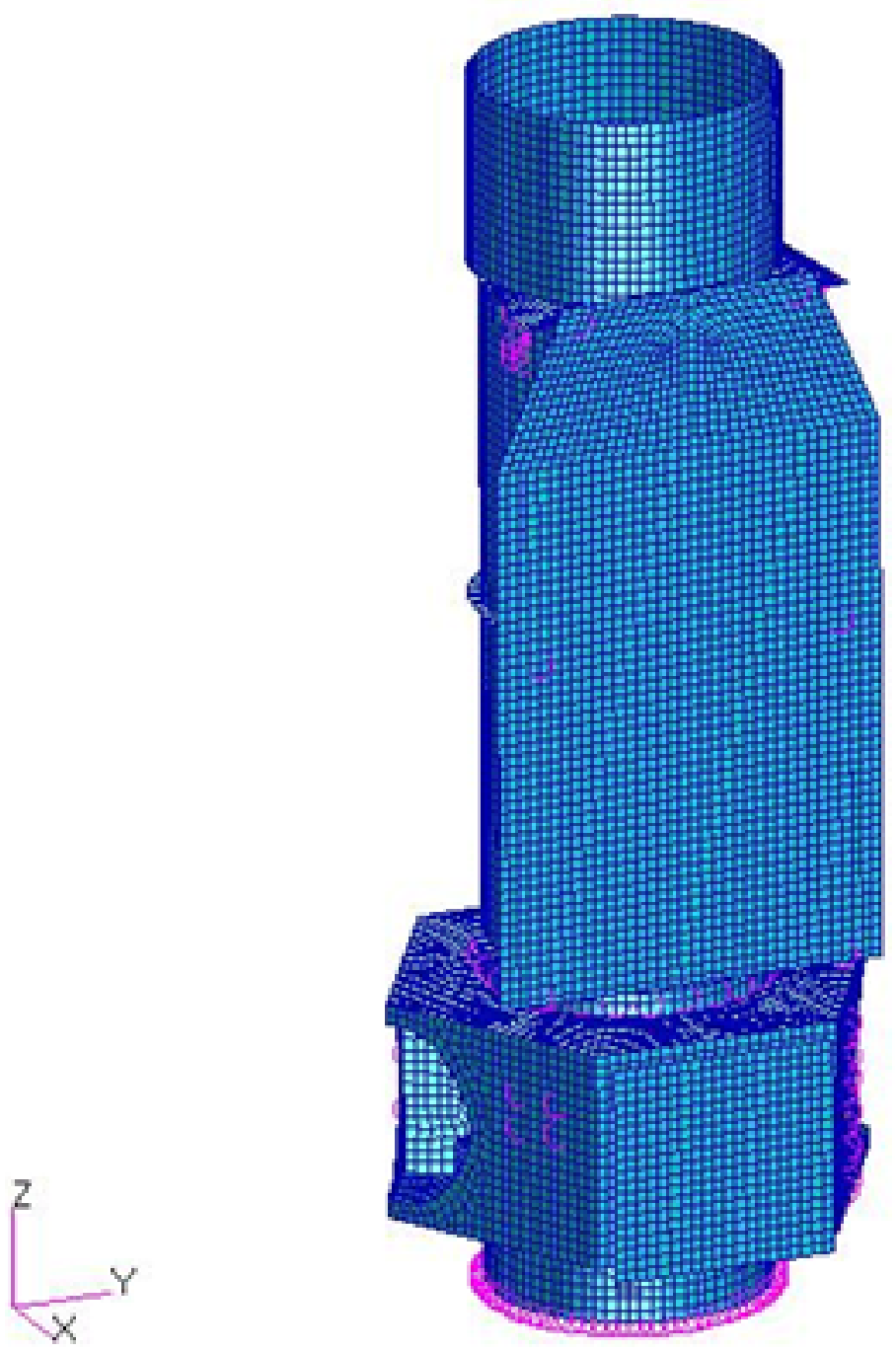}}
\caption{({\bf a}). POLARIX FEM with coordinate system({\bf b}).
POLARIX FEM with cooerdinate system}
\end{figure}

\subsection{The Satellite pointing and attitude control}

POLARIX satellite is a  3-axes attitude controlled platform able
to perform the fine pointing and the slew manouvre requested by
the scientific mission. The recognized design drivers for
POLARIX pointing required to meet the scientific mission goal are  shown in
table \ref{tab:6}. Attitude Measurement Error (AME) and Absolute
Pointing Error (APE) have been considered for each axis normal to
the spacecraft Line of Sight (LOS). AME and APE are considered at
95$\%$ confidence level (2 sigma  with normal distribution error)
on the set of all the possible scientific observations (spacecraft
pointing).

For  satellite  re-pointing (up to two times each orbit)  the
angular acceleration to be considered is greater than
$0.007deg/s^{2}$ with  an angular rate up to $0.8 deg/s^{2}$.

During the POLARIX phase A study,  a preliminary design of Attitude
Control Subsystem has been developed, demonstrating that by a
proper selection of on-the-shelf sensors/actuators,  the pointing
and attitude control of the satellite can be met by a proper
selection of  standard/off-the-shelf  equipment.

For POLARIX pointing and attitude determination and control it is
relevant to highlight the identification of  two engineering
solutions, with advantages in terms of cost and mission reliability,
i.e.: -   attitude determination approach  full gyroless by
means of  2 or 3 optical heads; -   the adaptations and/or minor
customizations of  Attitude and Control software developed in
others projects.

\begin{table}
\caption{POLARIX pointing design requirements}
\label{tab:6}       
\begin{tabular}{lll}
\hline\noalign{\smallskip}
Description & Mandatory Requirement or constraint & nice to have \\
\noalign{\smallskip}\hline\noalign{\smallskip}
Absolute measurement accuracy & 10'' 5 Hz  &  10'' 10 Hz\\
Absolute pointing error &  5' & 3'\\
Total number of pointing  & 150$/$yr & 2$/$orbit\\
Sky accessible  & 90$^\circ$$\pm$20$^\circ$ & 90$^\circ$$\pm$30$^\circ$\\
forbidden direction & none & none\\
\noalign{\smallskip}\hline
\end{tabular}
\end{table}

\subsection{Mass and power budget}
The summary mass budget is in table \ref{tab:10}. The power budget
of the entire satellite is in table \ref{tab:11}.

\begin{table}[!hhh]
\caption{Summary mass budget}
\label{tab:10}       
\centering
\begin{tabular}{ll} \hline\noalign{\smallskip}
Item & weight \\
\noalign{\smallskip}\hline\noalign{\smallskip}
Total Service Module Mass (kg) & 389  \\
Total Payload Module Mass (kg) & 335 \\
System level margin (20$\%$) & 145\\
Launcher adapter  &  45 \\
Total Mass at launch (kg) &  913 \\
\noalign{\smallskip}\hline
\end{tabular}
\end{table}

The  total mass of  913 Kg allows a very large margin for  launch
with Vega (capability of 2300 kg  for injection at the selected
orbit).

\begin{table}[!hhh]
\caption{Power budget}
\label{tab:11}       
\begin{tabular}{lllllll}
\hline\noalign{\smallskip}
S/S   & LEOP    & Sunlit Contact (W) & Eclipse Contact (W)  & Sunlit NoCont & Eclipse NoCont & Remarks\\
\noalign{\smallskip}\hline\noalign{\smallskip}
DHS  & 32.4 & 32.4 & 32.4 & 32.4 & 32.4 &  \\
AOCS  & 14.0 & 169.2  & 169.2 & 169.2 & 169.2 &  \\
PWR   & 44.0 & 644.0  & 44.0  & 644.0  & 44.0  & \\
TT$\&$C  & 16.8 & 32.4 & 32.4 &  16.8 & 16.8 & \\
P$\/$L & 0.0 & 56.4 & 56.4 & 56.4 & 56.4 & \\
TCS  & 0.0 & 120.0 & 120.0 & 120.0 & 120.0 & \\
S$/$S Total & 107.2 & 1054.4 & 454.4 & 1038.8 & 438.8 & Sun Ecl\\
Harness Losses & 3.2 & 13.6 & 13.6 & 13.2 & 13.2 & 3$\%$\\
PCDU losses   & 6.4 & 63.3 & 9.1 & 62.3 & 8.8 & 6$\%$   2$\%$\\
System Margin    & 21.4 & 66.9   & 66.9   & 63.8    & 63.8   20$\%$\\
Total  & 138.3 & 1198.2 & 544.0 & 1178.1 & 524.5 & \\
\noalign{\smallskip}\hline
\end{tabular}
\end{table}

\section{Ground Segment and User Segment}
The POLARIX mission is expected to operate in a low-Earth
equatorial orbit, using the ASI Ground Station in Malindi (Kenya)
as a primary ground station. The onboard telemetry production rate
is limited to 50 kbit/s. The satellite mass memory will be fully
downloaded once per orbit, when the satellite passes over Malindi.
The total 300 Mbit of each data download will be transmitted,
through the ASINET link, to the Telespazio facility in Fucino. The
ground support of the POLARIX mission will consist of two main
elements: the Ground Segment (G/S) and the User Segment (U/S). The Ground
Segment  will perform two main tasks: operate the POLARIX
satellite during all the phases of the mission in nominal and
contingency conditions; serve the scientific observation
requests during the operational phase of the mission. The POLARIX
G/S performs the main functions/operations needed at
ground level to manage the mission both in terms of satellite
control and global data management. The planned G/S includes the
ground station of the Italian Space Agency (ASI) located at
Malindi and the Mission Operation Center (MOC) located at the
Telespazio Fucino Space Center. The G/S provides all the functions to:
\textit{(a)} monitor and control the satellite
platform and payload; \textit{(b)} perform the orbit/attitude
operations and generation of orbit products used within the G/S
for satellite M$\&$C, payload management, mission planning and
product generation; \textit{(c)} generate the mission planning
and checks, according to scientific observation requests coming
from the U/S; \textit{(d)} acquire the raw satellite data
(housekeeping and telemetry) and transfer them to the U/S for
processing. The G/S includes also the Satellite Simulator and the
Communication Network responsible to interconnect the Ground
Segment facilities and to provide the related communication
services in a secure and reliable way.

\begin{figure}
\centering
\includegraphics[scale=0.5] {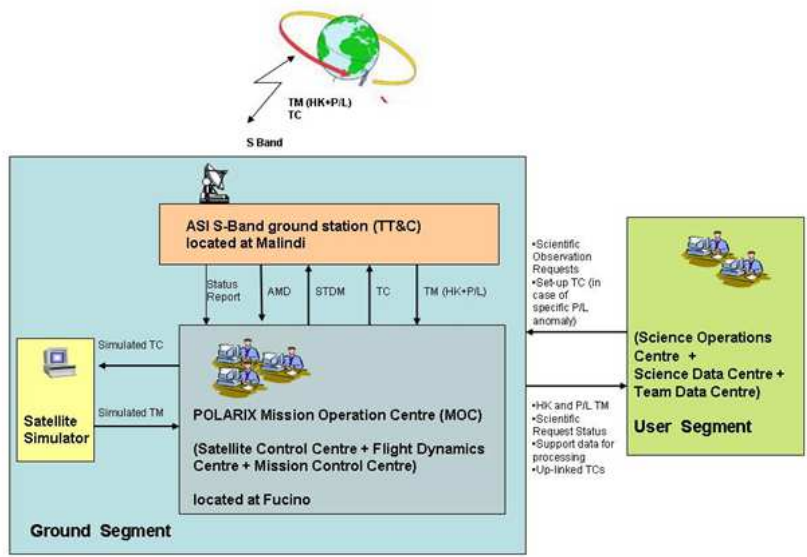}
\caption{POLARIX Ground and USER Segment}
\label{fig:43}       
\end{figure}

The U/S manages the scientific
observation requests coming from the scientific community,
forwards them to the G/S for payload scheduling activities,
ingests the raw satellite data coming from the G/S, generates,
archives, catalogues and delivers the scientific data and the data
products to the user community. Communication between the Fucino
Mission Operations Centre  and the Malindi ground station will be
allowed through the ASINET infrastructure provided by ASI. A
further connection between the MOC and the scientific community
will be possible by means an ISDN line which will ensure the
communication between the ASINET node in Fucino and the User
Segment located in Frascati.

\subsection{Data policy}
POLARIX is a PI mission but since its scientific exploitation is
open to the worldwide community, it will be conducted through a
standard observatory management. The scientific objectives of the
POLARIX mission are expected to be achieved by pointed
observations typically lasting from one day to one week. The
POLARIX scientific program will be then composed of a sequence of
staring observations. A Core Program for the mission is planned in
the first year, covering 25$\%$ of the available net observing
time, ensuring that the main scientific goals of the mission are
achieved. The POLARIX Team will have the responsibility for the
scientific exploitation of the Core Program data. The remaining
75$\%$ of the observing time is planned to be accessible to the
world wide community on a competitive basis through an
Announcement of Opportunity to be issued before the launch. Both
Core Program and Guest Observer Program (GOP) can include Target
of Opportunity observations. After ascertaining the feasibility of
the observation, the proposals will be selected on the basis of
their scientific quality, as assessed at a peer review by a Time
Allocation Committee. A proprietary period of one year of data is
guaranteed, starting with the delivery of the data and the
relevant analysis software and calibration files. After one year
the data will be put in an open access archive. Guests will be
supplied with a Guest Observer Handbook to prepare proposals and
analyse data, with documented software tools, based on open source
codes and with all data needed for a full exploitation. The
distributed data will be based on lists of qualified events
including absorption point and time, energy and polarization
angle, plus the data on coverage, time windows and dead time. All
the data, including those of the GOP, will undergo a Standard
Scientific Analysis (SSA), aimed only at guaranteeing a high scientific
quality of the data, before their delivery to the observer
together with the results and products of the SSA. The analysis of
the tracks, the analysis of calibrations and the production of the
response matrices of the instrument are the responsibility of the
POLARIX Team.

\subsection{Data Analysis}
The software required to reduce and analyze the POLARIX data will
be developed by the POLARIX Team, who will take the responsibility
for distributing it to the Guest Observers, to document and to
maintain it. The POLARIX software will be based on license-free
programming languages. The POLARIX Scientific software will enable
the following analysis. The data reduction will start with the
pre-processing, that is the conversion of telemetry files into
separated FITS files, depending on the information they contain.
Output data from the pre-processing will be provided to the
scientific and quick look analysis. These packages will run
automatically after every satellite contact with the Malindi
ground station. A SSA is planned to
be carried out on all data as early as possible after the
completion of a continuous observation of a given target,
regardless of the data rights, with the aim of verifying and
guaranteeing the scientific quality of the data. The results of
the SSA will pass through a pre-defined set of quality checks, as
automated as possible. After successful verification, the results
of the SSA will be archived and made available to the owner of
the data rights through a password-secured Web page for
scientific exploitation, and to the POLARIX Team for Quick Look
Analysis and calibration purposes only. After the period of
proprietary data rights has expired, the SSA will be made publicly
accessible through the POLARIX Public Data Archive. A Quick-Look
Analysis (QLA) of the POLARIX data will be carried out under the
responsibility of the POLARIX Team on different time scales:
orbital, daily-incremental, and complete observation. The aim of
the QLA is to discover unusual and noticeable astrophysical
phenomena that are of interest to the scientific community and
require a prompt information distribution in order to carry out
related and/or follow-up observations. The software package for
carrying out the QLA will be developed and maintained by the
POLARIX Team. The processing-log and the results of the QLA will
be archived in a database. After passing all the standard data
processing (pre-processing, data reduction, standard analysis,
QLA) the POLARIX data will be delivered to the data rights owner
for that specific observation. He/she will receive the complete
set of reduced data (photon list and auxiliary data) needed for an
optimal scientific exploitation, together with the results of the
SSA. In order to carry out specific, non-standard analysis, or to
verify the results of the SSA, the POLARIX Team will distribute to
the GO a documented software package, that the GO will use under
her/his own scientific responsibility, the POLARIX Off-line
Scientific Analysis Software (POSAS). The POSAS will be developed,
maintained and documented under the responsibility of the POLARIX
Team, and new releases will be distributed periodically, together
with the relevant calibration files. The POSAS software will be
based on the FTOOLS software.

\subsection{Data Storage}
All the data and products at their different steps (Telemetry,
Level 1, Photon List, Products) will be stored and archived for
future access and use. All the archives will be accessible by
means of user-friendly web interfaces and will be open to the
individual GOs as their proprietary data, and to the general user
as the public data. Similarly, the calibration and the scientific
software needed for analyzing the accessible data will be archived
and made accessible by an open web interface. The data and the
CALDB will be written and stored in the OGIP-FITS standard format
and structure.
\section{POLARIX performances}
\subsection{Characteristics of the polarimeter}
The characteristics of the polarimeter are shown in table
\ref{tab:13}

\begin{table}
\caption{Characteristics of the polarimeter}
\label{tab:13}       
\begin{tabular}{ll}
\hline\noalign{\smallskip}
Characteristic & Value\\
\noalign{\smallskip}\hline\noalign{\smallskip}
window thickness & 50 $\mu$m \\
 pixel-size & 50$\mu$m   \\
 number of pixels & 105600 \\
 area of the GPD & 1.5 cm $\times$ 1.5 cm\\
 Focal Length & 3.5 m \\
 no. of telescopes & 3 \\
 Field of View & 15' $\times$ 15' \\
 Baseline mixture & He-DME (20$\%$-80$\%$) \\
 Dead time & 50 $\mu$s\\
 Time resolution & 8 $\mu$s\\
 Time accuracy & 2 $\mu$s \\
 Energy band & 2-10 keV \\
 Energy resolution & 20 $\%$ @ 5.89 keV\\
 Crab  total  counting  rate & 145.6 cnts/s/cm\\
 Crab MDP (10$^{5}$s) & 0.396 $\%$\\
 Sensitivity  & 12 $\%$ @ 1 mCrab in 10$^{5}$s\\
\noalign{\smallskip}\hline
\end{tabular}
\end{table}

\subsection{Polarimetric performances}

In the Gas Pixel Detector, filled with the proper mixture
of gas, the tracks produced by a
photoelectron in the gas can be visualized. From the analysis of the
tracks, the impact point of the photon and the angular
direction of the photoelectron are derived.

\subsubsection{Capability to measure the
    polarization of a cosmic source}.

In practice, this means to find a modulation of
the histogram of the emission angles,$\phi$,
(the so-called modulation curve), which exceeds that
produced by random fluctuations at a predefined
level of probability. This is usually expressed as
the Minimum Detectable Polarization (MDP).

\begin{equation}
MDP=\frac{4.29}{\mu \times S} \times \sqrt{\frac{S+B}{T}}
\end{equation}

where $\mu$ is the modulation factor, a number within 0 and 1, that corresponds
to the amount of modulation induced by 100 $\%$ polarized source. If the
modulation curve of a 100 $\%$ polarized source is fitted with:

\begin{equation}
M(\phi) = A{_{100\%}}+C{_{100\%}}\cos^{2}(\phi-\phi_{0})
\end{equation}

then the modulation factor $\mu$ is :

\begin{equation}
  \mu = \frac{M_{max} - M_{min}}{M_{max} + M_{min}} = \frac{C_{100\%}}{C_{100\%} + 2A_{100\%}}
\end{equation}

B is the background count rate (residual background
and diffused X-ray background), T is the net observing time in seconds
while S is the source count rate which depends
on the quantum efficiency of the detector ($\varepsilon$ which in turn
depends on the gas mixture, the gas thickness and the window transmission),
on the telescope effective area and on the spectrum of the X-ray source.

The MDP should not be confused with the
measurement error (Weisskopf et al., 2009, 2010 \cite{Weisskopf2009},\cite{Weisskopf2010}).
The level of background in an imaging device at the
focus of a telescope is in any realistic case
negligible with respect to the counts from the source.
This means that the polarimetric sensitivity is
limited by the fluctuations of the unpolarized
fraction of the source itself and the equation reduces
to:

\begin{equation}
MDP = \frac{4.29}{\mu \sqrt{S}} \times \frac{1}{\sqrt{T}}
\end{equation}

for a 99$\%$ confidence level. The sensitivity to the
polarization angle is also connected to this parameter so that
the MDP is the synthetic parameter describing the
performance. The MDP depends linearly on the modulation
factor and on the square root of the product of the
collecting area of the telescope and the conversion
efficiency of the gas. The baseline mixture, He (20$\%$)
plus DME (80$\%$), is already compliant with the scientific
requirements, but the search for other mixtures or fine
tuning of the thickness of the absorption gap and/or of
the filling gas pressure, to further improve the
polarimetric sensitivity, will continue during the
development phases, since this will not impact on the
design and development of the other subsystems. The
practical parameter to compare the performances of
different polarimeters at the focus of the same telescope
is the quality factor (QF):

\begin{equation}
QF = \mu\sqrt{\varepsilon}
\end{equation}

\begin{figure}[htpb]
\centering
\subfigure[\label{fig:41a}]{\includegraphics[scale=0.25, angle=90]{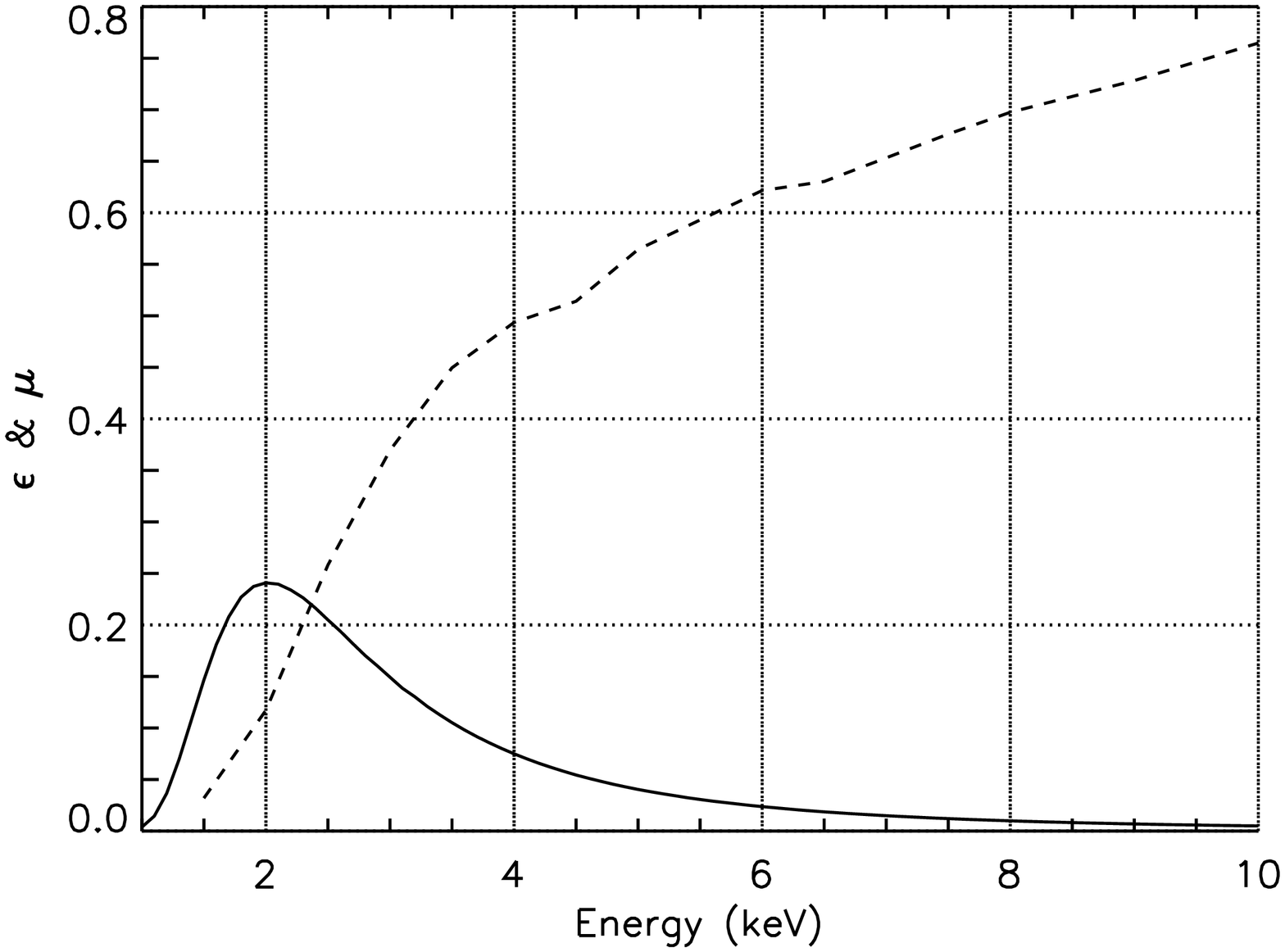}}
\hspace{0.2cm}
\subfigure[\label{fig:41b}]{\includegraphics[scale=0.25,angle=90]{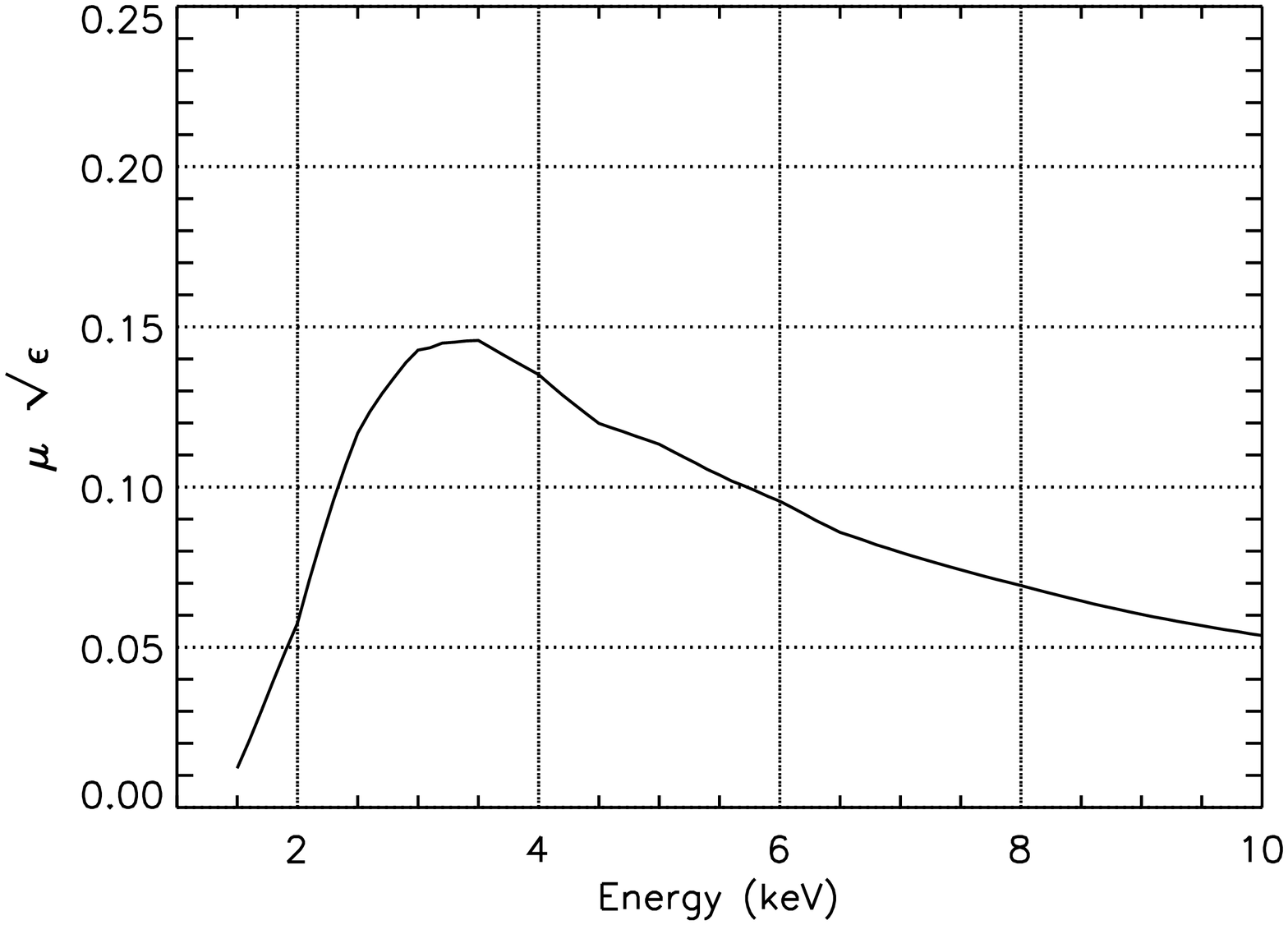}}
\caption{({\bf a}). Modulation factor (dashed line, $\mu$) as a function of energy
and efficiency (solid line, $\varepsilon$, included the window transmission), for the baseline mixture (20$\%$He-80$\%$DME) ({\bf b}).
Quality factor as a function of energy}
\end{figure}
The modulation factor ($\mu$), the efficiency ($\varepsilon$) and the quality
factor are shown in fig. \ref{fig:41a} and in fig.
\ref{fig:41b}. The modulation factor increases with energy since the emission angle determination is
more accurate for longer tracks, while the efficiency drops at low energies because of the window transmission (50 $\mu$m of Beryllium)
and at high energies due to the decreasing opacity of the gas mixture.

The Minimum detectable polarization for the POLARIX
mission is shown in fig.\ref{fig:42}

\subsubsection{Capability to control systematic effects}

    Control of systematic effects
    has been and still is a dramatic
    limitation for polarimeters based on Compton
    scattering. On the other hand, the complete visualization of the track
    and the intrinsic imaging capability of the GPD,
    makes this device free from any major systematic
    effects. In fact we were not able to detect any
    significant spurious modulation on signals detected
    from unpolarized photons, down to 1$\%$ level. An
    accurate study of any possible source of systematics
    will be in any case performed
    during the whole development and calibration activity.
    This will include effects due to the telescopes, or
   deriving from disuniformities of the detector, or
    from the track analysis algorithms.

\begin{figure}
\centering
\includegraphics[scale=0.5,angle=90] {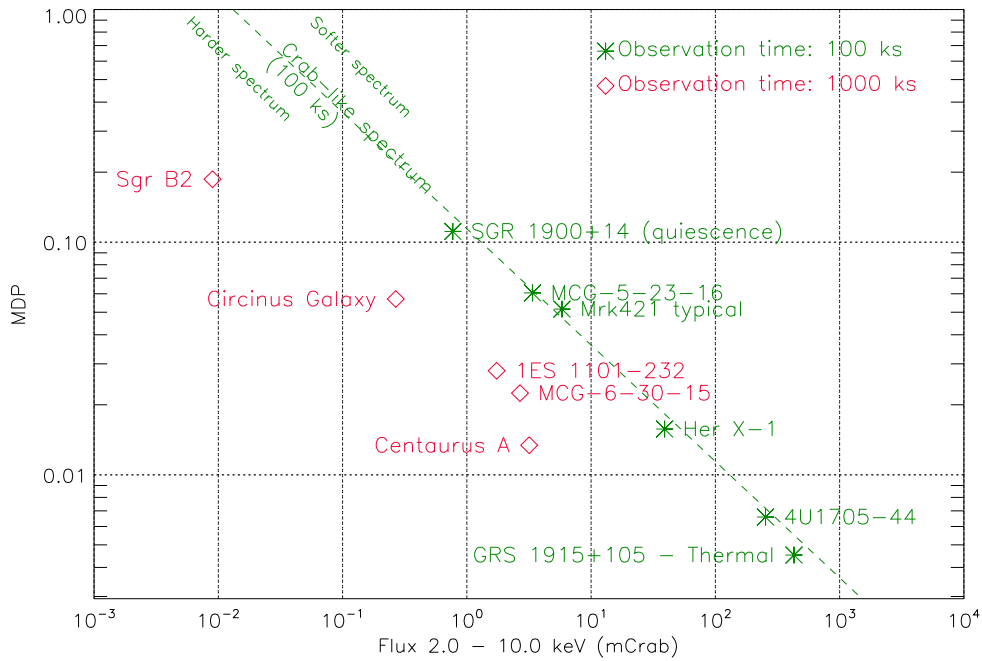}
\caption{Minimum detectable polarization for POLARIX}
\label{fig:42}       
\end{figure}

\subsection{Imaging performances}
Imaging capability is very important for three main reasons:
\begin{itemize}

\item To reduce the background

\item To single out the
    target source from others sources in the f.o.v..

\item  To perform
    angularly resolved polarimetry of extended
    sources (e.g. Pulsar Wind Nebulae).

\end{itemize}

The imaging capabilities of POLARIX are widely predefined from
already existing items. The telescopes of JET-X have an angular
resolution of about 15''. The on-axis imaging capability of
the GPD should be compliant with this performance, but, in
practice, it will be spoiled by the thickness of the absorption
gap, combined with inclination of X-rays from the telescope. The
total effect is of the order of 27''. If more telescopes
are added to POLARIX, manufactured from the JET-X mandrels, the
quality of the shells can be relaxed in a trade-off of weight and
resolution, with the goal to preserve a resolution of at least 40'' also
for these new telescopes.

\subsection{Spectroscopic performances}
Narrow lines are expected in most cases to be
unpolarized. Therefore a high spectral resolution is not required
for the science of POLARIX. A moderate energy resolution of the
order of 30$\%$ could be suitable to perform energy resolved
polarimetry of source continua. Nevertheless, since the modulation
factor and the efficiency are a relatively fast function of the
energy, better energy resolution would help to disentangle the
energy and polarimetric information. With the energy resolution
provided by the GEM we hope to reach a resolution of at least
20$\%$ (@6keV). In fig.\ref{spettro}) a spectrum (2.6 keV,
5.2 keV and 7.8 keV), obtained by using our calibration facility,
is shown (\cite{Muleri2010},\cite{Muleri2008b})

\begin{figure}
\centering
\includegraphics[scale=0.3] {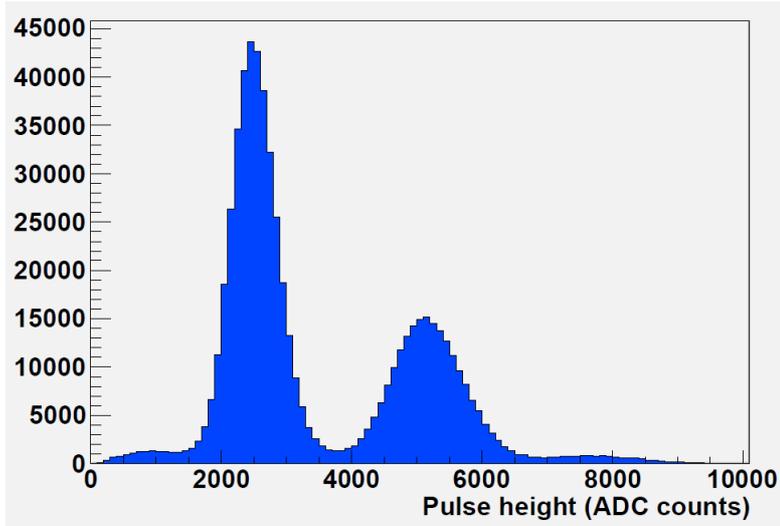}
\caption{Spectrum of the first three order of
diffraction (2.6 keV, 5.2 keV, 7.8 keV) of X-rays from a Graphite crystal collected by a GPD
filled with 0.8 bar of DME (Muleri et al 2010 \cite{Muleri2010})}
\label{spettro}       
\end{figure}

\subsection{Timing performances}
The signal from the GEM is intrinsically very fast. The timing
capability is limited by the shaping time of the electronic
chains, required for low noise. This is still far superior than
any scientific requirement. We fixed a timing resolution of 8
$\mu$s that preserves a large margin of discovery. In order to
synchronize the UT with OBT, POLARIX will have a GPS which allows
a synchronization at the level of 1-2 $\mu$s.

%
%
\section{Observing capability} \label{sec:observing}

POLARIX is a mission totally dedicated to X-ray polarimetry. This
important sub-topics of X-ray Astronomy, is always performed in
the 'source dominated' regime, given the large number of source
counts necessary to be collected for a sensitive measurement. The
MDP, in the 2-10 keV energy
band, with $10^{5}$ s of observation is 3 $\%$ for a source of 10
mCrab and it scales down simply with $\sqrt{Time \times Counts}$.
It is therefore possible to increase the sensitivity by observing
with a long pointing (one week/10 days) when required by dim
sources. In the following we provide estimates of the MDP with
reasonable observing time for different classes of celestial
sources.

\begin{itemize}

\item

    Magnetars are expected to be highly polarized for their
    large magnetic field. Bright magnetars, like SGR1900+14
    (MDP of 10$\%$ in 100 ks when in quiescence) can be looked
    at to start searching for QED effects and other very
    interesting effects, as for instance the presence of
    axions (Lai \& Heyl 2007 \cite{Lai2006}), the elusive
    particles which are candidates for the main dark matter
    component.

\item

    The nebulae around $SgrA^{\ast}$, namely Sgr B2 and Sgr C,
    could  be, actually,  studied with a 10 days long pointing:
    X-ray polarimetry with POLARIX thanks to its rather good
    spacial resolution would solve the puzzle of the origin of
    their X-ray emission most probably due to past activity of
    the central galaxy black hole. In the case of  Sgr B2 a
    MDP of about $20\%$ can be obtained with 1 Ms observation.

\item

 The angular resolution and the sensitivity of POLARIX would
 allow us to perform a spatially resolved X-ray polarimetry of the
 prototype of the PWNe, the Crab Nebula and Pulsar. The FoV of
 POLARIX allows us to perform imaging polarimetry of Crab Nebula
 within a single observation. It would then be possible to
 separate the torus emission from the jets integrating the
 X-ray polarization in a few independent regions. X-ray
 polarimetry would be useful to start to clarify the role of
 the magnetic field and of the particle flow in PWNe.  Vela
 pulsar region which shows similarity with the Crab nebula
 could be also studied and X-ray polarimetry would be
 performed in two independent regions.

\item

 X-ray polarimetry of thermal emission from Microquasars
    hosting a Black Hole derived from the accretion disk can
be performed in a very detailed way by  POLARIX. The X-ray
position angle and degree of polarization could therefore be
studied as a function of energy providing information on the
spinning of the black hole. Actually the best source to search
for this effect is indeed GRS1915+105 (see figure
\ref{fig:11a} and \ref{fig:11b}), a bright Microquasar whose
2-10 keV emission is, when in high state, dominated by thermal
emission. Moreover, the source is highly inclined (70 degrees,
Mirabel $\&$ Rodriguez 1994 \cite{Mirabel1994}), and therefore the polarization
degree is expected to be high. GRS1915+105 can be studied at
level of 0.34 $\%$ in one day of observation. The possibility
to perform X-ray polarimetry simultaneously to radio emission
will allow the study of the interaction between the jet, the
corona and the accretion disk. Polarization from Cyg X-1,
Cyg-X3, J1655-40 and SS-433 could be studied in short interval
of time allowing variability studies. Sources like Cyg X-1
(black hole) and 4U1705-04 (neutron star) show clear evidence
for an accretion disc, they are bright enough to search for
polarization less than 1$\%$ in 100 ks.

\item Pulsators are provided with a magnetic field of the
    order of 10$^{12}$ Gauss. X-rays are therefore expected to
    be highly polarized especially at energy close to
    cyclotron lines. Energy resolved and phase resolved
    polarimetry can be performed by POLARIX at level of 1 $\%$
    in the first minumum and in the first part of the ascent
    for Vela X-1 the prototype of X-ray pulsators. Polarimetry
    not energy or phase resolved would fail to detect
    polarization because of the variation of the angular phase
    as was the case for OSO 8. As in the case of PWNe the
    swing of polarization angle with the angular phase will
    help to fix the angle between the magnetic and the
    rotation axis. Energy dependent polarimetry with POLARIX
    can confirm or disprove this scenario in a number of
    bright X-ray pulsars (e.g. Vela X-1, Her X-1, Cen X-3,
    A0535+26, etc.) by searching for polarization as low as a
    few percent in several phase bins and different energy
    bins.

\item Cataclismic Variable (CVs) can polarize the radiation
    either because of the reflection on the WD surface (Matt
et al., 2004 \cite{Matt2004}) of X-rays produced by the
shocked accreting column or because they are scattered by the
electrons which precipitate onto the White Dwarf (Wu 2010
\cite{Wu2010}). In the latter case for particular geometry the
polarization can be as high as 8 $\%$. Am Her  can be as
bright as 5 mCrab providing a MDP of about 5 $\%$ in $10^{5}$
s of observation. A smaller MDP can be obtained with a longer
pointing.

\item Regarding millisecond X-ray pulsar (MSP), POLARIX is sensitive
    enough to study its X-ray polarization. The quiescent
    emission of  MSPs is usually rather weak, but still
    allowing for meaningful measurements of polarization with
    POLARIX (e.g. MDP of 3.7 $\%$ in 100 ks for
    SAX1808.4-3658, which becomes 1.2$\%$ when in outburst).

\item In radio quiet AGN the corona play a crucial role in
    producing X-ray. The models can be tested with POLARIX at
    least for the brightest objects. Actually in the case of
    IC4329A, NGC5006 and MCG-5-23-16, MDP of a few percent can
    be reached in 100 ks, enough for a first test of the
    Comptonization model. In Compton-thick AGN, the reflection
    component dominates the 2-10 keV band. The brightest of
    such sources is the Circinus Galaxy (Matt et al.
    1996\cite{Matt1996}) for which a MDP of about  6$\%$ can
    be reached in a 1 Ms observation.

\item
 Blazars are among the most promising sources for X-ray
    polarimetry. Mkn421, for example, shows its X-ray emission in
    the synchrotron peak,  therefore a high polarization is
    expected. When in flare Mrk421 will be observed with a MDP
    of 1 $\%$ in $10^{5}$ s while in a typical state it will
    be observed with a MDP of $5\%$ . Multi-wavelength
    polarimetric observation could help to study the magnetic
    field and its microvariability and the energy distribution
    of the emitting particles. The case of 3C 454.3 is
    different, X-ray emission lies in the inverse Compton
    peak. The inverse Compton can be modelled either as
    Synchrotron Self Compton (the jet up-scatters soft
    synchrotron photons), or as External Compton from seed
    soft photon not produced in the jet scattered from the jet
    itself. POLARIX would provide a clue in disentangling
    between the two models and provide a clue about the physical
    status of the electron in the jet and may be providing an
    insight on the origin of the soft photons. POLARIX will be
    able to detect X-ray polarization at level of 3 $\%$ in 10
    day of observation from 3C 454.3. Many blazars are bright
    enough to allow a significant polarization measurement
    with POLARIX (MDP of a few percent in 100 ks). In non
    Blazar radio galaxy at least two sources, Centarus A and
    3C 273, are bright enough to perform an energy-dependent
    polarimetry with POLARIX down to a few percent in a  few
    days exposure.

\end{itemize}

    More exotic physics can be studied with POLARIX. Very stringent
upper limits on Quantum Gravity in the loop representation  can be
derived by observing distant blazars. With an observation of
10$^{6}$ s, values of $\eta$ down to 3$\times$10$^{-10}$ can be
measured with POLARIX using e.g. the known Blazar 1ES1101+232, at
z=0.186, with clear synchrotron spectrum and high optical
polarization, assuming it has a 10$\%$ polarization degree in the
X-ray band.

    We are confident that the number of celestial sources in each class
with measurable polarization angle and degree discussed above for
a pathfinder mission will adequately fill the planned  one year
of satellite operation in orbit. However the sensitivity of
POLARIX would allow the extension of this allocated time for
sensitive measurement of sources within each class.

\begin{acknowledgements}

The first studies of POLARIX were supported by ASI contract
I/088/060 for which we thank Elisabetta Cavazzuti. The phase A
study was supported by ASI contract I/016/08/0 . We thank Maria
Cristina Falvella, Donatella Frangipane, Elisabetta Tommasi,
Simona Zoffoli, Marino Crisconio, Paolo Giommi, Francesco Longo,
Fabio D'Amico, Jean Sabbagh, Giancarlo Varacalli and Valeria
Catalano from ASI for the effective and cooperative effort during
all the study. We also acknowledge the support of ASI contract
I/012/08/0 and Maria Barbara Negri. We also acknowledge the
contrbution of large teams of Thales Alenia Space Italy and of
Telespazio. A special thanks to Volker Liebig, Director of Earth
Observation Programmes of ESA, for considering the possibility of
making available to POLARIX spare parts of the GOCE mission.

\end{acknowledgements}

\bibliographystyle{spmpsci}      
\bibliography{References}   

\end{document}